\documentclass{article}

\usepackage{arxiv} 
\usepackage[utf8]{inputenc} 
\usepackage[T1]{fontenc} 
\usepackage{authblk} 
\usepackage{authblk}
\usepackage{microtype} 
\usepackage{xcolor} 
\usepackage{hyperref} 
\usepackage{url} 

\usepackage{graphicx} 
\usepackage{booktabs} 
\usepackage{tabularx} 
\usepackage{threeparttable} 
\usepackage{float} 
\usepackage{multirow}

\usepackage{pgfplots}
\usepackage{tikz}
\usepackage[labelfont=bf]{caption}
\usepackage[most]{tcolorbox}

\usepackage[table]{xcolor}
\usepackage{colortbl}
\usepackage{booktabs}

\usepackage{amsmath}
\usepackage{amssymb}

\usepgfplotslibrary{
  groupplots,      
  statistics,      
  colorbrewer,     
  fillbetween,     
  dateplot         
}
\usetikzlibrary{patterns, arrows.meta, positioning, calc, fit}

\usepackage{pgfplotstable} 

\usepackage{filecontents} 

\usepackage[labelfont=bf]{caption} 
\usepackage{subcaption} 

\usepackage{siunitx}
\usepackage{enumitem} 

\definecolor{myblue}{RGB}{98,139,180}
\definecolor{myorange}{RGB}{246,153,63}
\definecolor{mygreen}{RGB}{120,194,173}
\definecolor{mypurple}{RGB}{173,120,194}
\definecolor{softblue}{RGB}{98,139,180}
\definecolor{softpink}{RGB}{238,126,130}

\pgfplotsset{
  every axis/.append style={
    tick label style={font=\small},
    label style={font=\small},
    title style={font=\small},
    legend style={font=\small},
    grid=both,
    grid style={line width=.2pt, draw=black!10},
    major grid style={line width=.3pt, draw=black!15}
  },
  cycle list/Dark2-6, 
}

\usepackage{fancyhdr}
\begin{document}


\title{Generative AI in Saudi Arabia: A National Survey of Adoption, Risks, and Public Perceptions}

\date{} 

\renewcommand{\headrulewidth}{1pt}
\renewcommand{\footrulewidth}{0pt}
\fancyhead{}
\fancyfoot{}


\vspace{-6em}

\author[1]{Abdulaziz AlDakheel}
\author[1]{Ali Alshehre}
\author[1]{Esraa Alamoudi}
\author[1]{Moslim AlKhabbaz}
\author[1]{Ahmed Aljohani}
\author[1]{Raed Alharbi}

\affil[1]{College of Computing and Informatics, Saudi Electronic University, Riyadh 11673, Saudi Arabia}

\renewcommand{\shorttitle}{\textit{arXiv} Template}

\hypersetup{
pdftitle={A template for the arxiv style},
pdfsubject={q-bio.NC, q-bio.QM},
pdfauthor={David S.~Hippocampus, Elias D.~Striatum},
pdfkeywords={First keyword, Second keyword, More},
}

\hypersetup{
    colorlinks=true,
    urlcolor=blue,
    linkcolor=blue,
    citecolor=blue
}

\maketitle

\begin{abstract}

Generative Artificial Intelligence (GenAI) is rapidly becoming embedded in Saudi Arabia’s digital transformation under Vision 2030, yet public awareness, adoption, and concerns surrounding these tools remain unexplored. This study provides an early snapshot of GenAI engagement among Saudi nationals. Using a nationwide survey of 330 participants across regions, age groups, and employment sectors, we examine seven dimensions of GenAI use: awareness and understanding, adoption patterns, perceived impacts, training needs, risks and barriers, data-sharing behaviors, and future expectations. Findings show that 93\% of respondents actively use GenAI primarily for text-based tasks while more advanced uses such as programming or multimodal generation are less common. Despite the prevalence of use, overall awareness and conceptual understanding remain uneven, with a majority reporting limited technical knowledge. Participants recognize GenAI’s benefits for productivity, work quality, and grasping complex information, yet many caution that sustained reliance on these tools may undermine critical thinking and key professional skills. Trust in AI-generated outputs remains cautious, with widespread concerns about privacy, misinformation, and ethical misuse, especially in relation to potential job displacement. Respondents show strong interest in structured GenAI training that combines foundational skills, domain-specific applications, and clear guidelines on privacy, ethics, and responsible use. These results establish a foundational baseline for GenAI engagement in Saudi Arabia and highlight three priorities for policymakers and developers: expanding AI literacy, ensuring culturally and linguistically aligned GenAI solutions, and strengthening frameworks for privacy and responsible deployment.

\end{abstract}

\section{Introduction}
\label{sec:Introduction}
\begingroup

GenAI is rapidly reshaping the current technological landscape, redefining how individuals create, communicate, and interact with information. As global attention shifts toward understanding and regulating this technology, the Kingdom of Saudi Arabia is pursuing a comprehensive digital transformation under its national roadmap, Vision~2030. This strategy aims to diversify the economy, reduce dependency on oil, and position the country as a regional leader in technology and innovation. Within this context, GenAI plays a central role in driving progress across education, healthcare, government services, and creative industries.

Building on this vision, the Saudi Data and Artificial Intelligence Authority (SDAIA) leads national efforts to ensure the responsible and ethical development of AI. The Kingdom’s central authority defines regulatory frameworks, guides implementation through its specialized centers, and ensures that innovation aligns with cultural and societal values. In its 2024 public report~\cite{sdaia_report}, SDAIA projects global GenAI spending to reach USD 202 billion by 2028, with productivity gains up to 30\% across sectors. SDAIA has also issued national \textit{Generative AI Guidelines}, emphasizing responsible design, human oversight, and transparency~\cite{sdaia_guidelines}.

Complementing this regulatory foundation, HUMAIN\footnote{\url{https://www.humain.ai/en/}}
 serves as Saudi Arabia’s innovation and infrastructure engine, building the computational backbone for large-scale AI development. In collaboration with global leaders such as NVIDIA and AMD, HUMAIN is building multi-gigawatt data centers, sovereign cloud platforms, and high-performance GPU clusters. These infrastructures will power Arabic large language models (LLMs) and enable human-centered applications, customized solutions, and exportable AI capabilities. These efforts support the Kingdom’s ambition to emerge as a global leader in artificial intelligence.

Despite this rapid progress, a critical research gap persists in understanding how Saudi nationals perceive, adopt, and build trust in GenAI. Numerous international studies examine awareness, ethical perceptions, and adoption behaviors in Western, East Asian, and broader global contexts. However, empirical evidence from the Gulf region remains rare. Saudi Arabia’s sociocultural and linguistic distinctiveness where Arabic is the primary language of digital interaction and cultural norms strongly influence technology adoption combined with a predominantly young and digitally native population, makes it difficult to generalize findings from other regions. In addition, the Kingdom’s regulatory and ethical frameworks, which emphasize responsible innovation, data protection, and cultural integrity, introduce further dimensions that shape user trust and acceptance in ways not yet examined in existing work.

Understanding how nationals perceive the opportunities and risks of GenAI is therefore not only an academic concern but also a national priority. Insight into public attitudes toward automation, privacy, misinformation, and cultural representation is essential for evidence-based policymaking, ethical AI governance, and the design of public-facing GenAI services. Such insight can help ensure that AI adoption remains aligned with the strategic objectives of Vision~2030 and that rapid technological change is matched by societal readiness, transparency, and trust. It also has direct implications for education, industry, and the public sector, where GenAI tools are already starting to shape daily practices.

To address this gap, this study conducts a nationwide survey of 330 Saudi nationals representing diverse age groups, sectors, and regions across the Kingdom. The survey explores awareness and understanding, adoption patterns, perceived impacts, training needs, risks and barriers, data-sharing behaviors, and future expectations regarding GenAI. Specifically, we investigate the following seven research questions (RQs):

\begin{itemize}

    \item \textbf{RQ1:} How frequently do different population segments use GenAI tools, and in which contexts (e.g., personal, educational, professional) and tasks do they employ them?

    \item \textbf{RQ2:} What is the current level of awareness and technical understanding of GenAI among nationals in Saudi Arabia?

    \item \textbf{RQ3:} How do users perceive the impact of GenAI on productivity, creativity, learning, decision-making, and core skills?

    \item \textbf{RQ4:} Which GenAI training topics are users most interested in, and which combinations of topics are most commonly selected together?

    \item \textbf{RQ5:} How do demographic and contextual factors (e.g., age, education, employment status, sector) influence perceived barriers, concerns, risk, and trust perceptions related to GenAI?

    \item \textbf{RQ6:} How does willingness to share personal data with GenAI tools vary by users status and data type?

    \item \textbf{RQ7:} What future expectations, policy preferences, and personal reflections do users express about the role of GenAI in their lives and in Saudi society more broadly?
\end{itemize}

By addressing these research questions, this study provides empirically grounded insights into how GenAI is perceived, adopted, and trusted within the Saudi context. The findings offer actionable guidance for policymakers, educators, and technology developers seeking to ensure that GenAI deployment remains aligned with the Kingdom’s cultural identity, ethical frameworks, and national innovation priorities under Vision~2030.

The remainder of this paper is organized as follows. Section \ref{sec:lite} reviews related efforts on GenAI adoption and digital transformation. Section \ref{sec:methodology} describes the survey design, data collection, and analysis methodology. Section \ref{sec:Results} presents the key quantitative and qualitative findings. Section \ref{sec:Limitations} discusses the limitations of the current study. Finally, Section \ref{sec:Conclusion} concludes the paper with the main insights and directions for future research.

\endgroup

\section{Literature Review}
\label{sec:lite}

GenAI adoption has remarkable speed, extending from broad public exposure to widespread student adoption and professional use. Since the release of ChatGPT in late 2022, it spreads at an extraordinary rate, exceeding the diffusion trajectories of earlier digital technologies such as the desktop internet \cite{suguna2021artificial}. By late 2024, nearly $40\%$ of U.S. adults use a GenAI tool \cite{bick-2024}, and similar patterns of adoption appear internationally. In higher education, adoption is especially widespread, with more than $90\%$ of undergraduates in the United Kingdom reporting engagement with AI tools - nearly $88\%$ using them in coursework \cite{freeman-2025}. Similarly, a global survey of over $23,000$ students shows that $71\%$ reported using ChatGPT within its first year \cite{ravselj-2025}. Beyond education, business communities also demonstrate rapid engagement, with two-thirds of leaders reporting GenAI use for both personal and organizational tasks \cite{davis2024early, syed2024awareness}.

While overall adoption rates highlight the remarkable diffusion of GenAI across populations and sectors, patterns of use are not uniform. Usage varies considerably depending on contextual and individual factors, including demographic profiles, professional roles, and disciplinary backgrounds. This emphasizes the importance of moving beyond aggregate statistics to examine various domain and subgroup differences that shape how GenAI tools are accessed, perceived, and integrated into daily practice.

Demographic factors strongly shape GenAI adoption and attitudes. Younger users show greater enthusiasm and higher usage than older cohorts, with business leaders under 30 significantly more engaged than those over 50 \cite{davis2024early}. Adoption also varies by education: STEM students and non-native English speakers use ChatGPT more frequently for coding and writing support, while humanities majors and native speakers engage less \cite{baek-2024}. Gender differences persist, as male students report higher usage and confidence than female peers, and male AI experts express greater optimism than women \cite{baek-2024, pew-2025}. Overall, age, education, field, and gender substantially influence adoption patterns, highlighting the need to consider subgroup differences across different domains.

The awareness of GenAI has expanded rapidly, laying the foundation for diverse patterns of adoption and use. By mid-2023, $58\%$ of US adults reported some level of familiarity with ChatGPT, whereas $42\%$ indicated no level of awareness \cite{pew-2025}. The highest levels of awareness are observed among younger and more educated groups \cite{pew-2025} and spread rapidly throughout academia. Building on this growing familiarity, adoption and usage span academic, professional, and personal domains, with patterns varying by context. Students most often use ChatGPT for summarization, brainstorming, and reference searching, while tasks such as creative writing or programming are less common \cite{ravselj-2025}. In professional settings, journalists rely on it for translation, transcription, and proofreading, whereas software engineers employ it for code generation and debugging \cite{d2025ai, wolf-2024}. Beyond these domains, everyday users integrate ChatGPT into routines such as travel planning, cooking guidance, and language learning \cite{wolf-2024}.

Perceived benefits and impacts constitute a central theme in emerging research, with studies consistently highlighting substantial gains in efficiency and productivity from GenAI use. Among business leaders, roughly $90\%$ identify time savings as the primary advantage, viewing GenAI as a tool to reduce routine workloads \cite{davis2024early}. Empirical evidence aligns with these perceptions: a six-month field study across multiple companies found that employees using an AI writing assistant reduced time on routine emails by $31\%$, saving $3.6$ hours per week and gaining nearly two additional hours for deep-focus work, without altering meeting loads \cite{dillon-2025}. 

In education, students report parallel benefits, describing GenAI as enhancing productivity and learning by simplifying complex information and offering on-demand explanations \cite{ravselj-2025}. Research on assignments further links ChatGPT use to perceived improvements in writing quality and comprehension. Among knowledge workers, GenAI appears to shift responsibilities from direct task execution to the oversight and integration of AI outputs, reducing cognitive load for routine work while increasing demands for verification and editing skills \cite{lee-2025}. Broader analyses reinforce the perception that GenAI improves efficiency and work quality, though the magnitude and nature of these effects vary considerably across contexts \cite{davis2024early, ravselj-2025}.

Alongside its reported benefits, GenAI has generated widespread concern, most notably regarding the accuracy and integrity of its outputs. Hallucinated and fabricated outputs represent a primary concern, with surveys showing that nearly two-thirds of both experts and the public fear GenAI disseminating false information \cite{pew-2025}. Additional risks include plagiarism, bias, and misinformation. Students often express anxiety about plagiarism accusations when incorporating AI into assignments \cite{ravselj-2025}, whereas journalists highlight ethical threats related to authenticity and the erosion of trust in news content \cite{d2025ai}. Privacy and data security also emerge as critical issues, with users reluctant to share sensitive information due to uncertainties over storage, misuse, and opaque handling practices \cite{alammari-2024}. At the organizational level, executives identify potential misuse as the foremost risk, ranking it even above concerns about job displacement \cite{davis2024early}. In the Saudi context, academics raise similar concerns, pointing to risks of over-reliance, biased or erroneous results, and threats to academic integrity \cite{syed2024awareness}.

Research on personal reflections and the future outlook highlight how individuals envision the trajectory of GenAI and the governance structures required to guide its use. Both lay users and experts consistently call for stronger oversight, with surveys indicating that more than half of US adults and AI specialists prefer greater personal control over AI integration \cite{pew-2025}. Approximately $60\%$ of both groups also express concern that government regulation is inadequate rather than excessive, and the majority report no confidence in companies or governments to regulate AI effectively \cite{pew-2025}. 

These concerns translate into sector-specific recommendations. In higher education, students advocate for clearer institutional policies and more extensive training: while 80\% of UK students considered their university's AI use policy clear, only 36\% reported receiving skills training, underscoring calls for better preparation for an 'AI-enabled' future \cite{freeman-2025}. Journalists emphasize the need for industry-wide ethical guidelines and detection tools to protect accuracy and transparency standards \cite{d2025ai}. Business leaders anticipate significant workplace transformations, and younger executives are particularly optimistic about leveraging GenAI for competitive advantage, while simultaneously stressing responsible adoption to mitigate risks \cite{davis2024early}. In Education, scholars look at AI similarly as a routine partner in teaching, and Saudi Arabia research recommends frameworks for responsible integration, including ethical training for instructors and students \cite{alammari-2024}.


Unfortunately, existing research on GenAI perceptions remains limited in various perspectives. Many studies focus on specific populations, such as university students \cite{pew-2025}, professionals in particular industries \cite{d2025ai}, or niche communities such as digital humanities scholars \cite{ma2025dancingbearcolleaguesharpened}. Although such studies provide useful insights, they fail to capture the diversity of experiences and expectations that shape the engagement with GenAI in broader societies, such as Saudi Arabia. 

In the Saudi context, research on GenAI perceptions remains minimal. To the best of our knowledge, no study has provided a comprehensive overview of public engagement across dimensions such as awareness, usage, perceived benefits, risks, and future outlook. This lack is particularly consequential given Saudi Arabia’s demographic profile as one of the world’s youngest and most digitally connected societies. Addressing this gap is crucial, as it not only captures local perspectives but also broadens global debates on the societal implications of GenAI by incorporating insights from a region that remains underrepresented in existing scholarship.

\label{sec:Literature}

\section{Methodology}
\label{sec:methodology}

\begin{table}[t]
\centering
\caption{Participant Demographics (N = 330)}
\label{tab:demographics}
\scriptsize
\begin{threeparttable}
\begin{tabular}{@{}p{2.1cm} p{3.3cm} r r@{}}
\toprule
\textbf{Category} & \textbf{Group} & \textbf{n} & \textbf{\%} \\
\midrule

\multirow{2}{=}{\raggedright Gender} 
    & Male & 172 & 52.12 \\
    & Female & 158 & 47.88 \\

\midrule
\multirow{7}{=}{\raggedright Age Group} 
    & 13--17 & 20 & 6.06 \\
    & 18--22 & 90 & 27.27 \\
    & 23--28 & 66 & 20.00 \\
    & 29--35 & 37 & 11.21 \\
    & 36--44 & 48 & 14.55 \\
    & 45--60 & 60 & 18.18 \\
    & 61+    & 9  & 2.73 \\

\midrule
\multirow{6}{=}{\raggedright Region} 
    & Central & 160 & 48.48 \\
    & Eastern & 87 & 26.36 \\
    & Western & 59 & 17.88 \\
    & Southern & 12 & 3.64 \\
    & Northern & 10 & 3.03 \\
    & Other / Unspecified & 2 & 0.61 \\

\midrule
\multirow{5}{=}{\raggedright Education Level} 
    & Bachelor & 216 & 65.45 \\
    & Other & 44 & 13.33 \\
    & Diploma & 33 & 10.00 \\
    & Master & 28 & 8.48 \\
    & Doctorate & 9 & 2.73 \\

\midrule
\multirow{5}{=}{\raggedright Employment Status} 
    & Student & 115 & 34.85 \\
    & Employee & 100 & 30.30 \\
    & Other & 45 & 13.64 \\
    & Student \& Employee & 35 & 10.61 \\
    & Unemployed & 35 & 10.61 \\

\midrule
\multirow{10}{=}{\raggedright Field of Study} 
    & Other & 167 & 50.61 \\
    & ICT & 88 & 26.67 \\
    & Education & 38 & 11.52 \\
    & Health \& Social Care & 16 & 4.85 \\
    & Arts \& Humanities & 8 & 2.42 \\
    & Data Science & 4 & 1.21 \\
    & Computer Science & 3 & 0.91 \\
    & Tourism & 3 & 0.91 \\
    & General Programs & 2 & 0.61 \\
    & Engineering & 1 & 0.30 \\

\midrule
\multirow{7}{=}{\raggedright Organization Size} 
    & Other / Unknown & 178 & 53.94 \\
    & Very Large (1000+) & 69 & 20.91 \\
    & Medium (50--249) & 26 & 7.88 \\
    & Small (6--49) & 23 & 6.97 \\
    & Large (250--1000) & 14 & 4.24 \\
    & Freelance / Self-employed & 10 & 3.03 \\
    & Micro (1--5) & 10 & 3.03 \\

\midrule
\multirow{9}{=}{\raggedright Sector} 
    & Other & 226 & 68.48 \\
    & Information Technology & 39 & 11.82 \\
    & Education & 35 & 10.61 \\
    & Government & 13 & 3.94 \\
    & Energy & 6 & 1.82 \\
    & Healthcare & 4 & 1.21 \\
    & Manufacturing & 4 & 1.21 \\
    & Construction & 2 & 0.61 \\
    & Real Estate & 1 & 0.30 \\

\bottomrule
\end{tabular}
\end{threeparttable}
\end{table}
\subsection{Study Design and Overview}
This study adopts a cross-sectional, mixed-methods design based on a nationwide survey of Saudi nationals. The survey examines awareness and technical understanding of GenAI, adoption and usage patterns, perceived impacts, training interests, risk perceptions, trust, and future expectations, aligned with the seven research questions (RQ1–RQ7). Quantitative analyses of closed-ended items are complemented by qualitative thematic analysis of open-ended responses.

\subsection{Population and Sampling}
The target population consists of Saudi nationals aged 13 years and above. We sought broad coverage across regions, age groups, educational levels, fields of study, and employment sectors within the Kingdom. The survey link was disseminated openly through online and offline channels, and participation was voluntary. Given that participation occurred through self-selection in these channels, the sampling approach is best described as non-probability convenience sampling with elements of snowballing rather than a strictly random sample.

In total, the survey collected $364$ valid submissions from the general public. We excluded responses from non-Saudi nationals and incomplete or low-quality submissions based on predefined data screening criteria. The final analytic sample consists of $330$ Saudi respondents, representing diverse age groups, sectors, and regions across the Kingdom.

Among these, 24 participants report not using GenAI tools. Their responses are excluded from analyses involving usage-dependent variables (e.g., daily usage frequency) to maintain analytical accuracy, but are retained in perception-based analyses, such as awareness and attitudes to provide broader insights into public understanding and exposure to GenAI. A descriptive overview of participant characteristics is presented in Table~\ref{tab:demographics}.

\subsection{Survey Instrument}
The survey is self-administered and consists of 31 questions. It was implemented in Microsoft Forms and combines multiple-choice, Likert-scale, and open-ended items covering six thematic domains: demographic characteristics, awareness and understanding levels, adoption and usage patterns, perceived impacts, Concerns and Risk Perceptions, and personal reflections.

To structure the questionnaire, we organized items into the following domains:

\paragraph{Demographics.}
We collect background information on age group, gender, nationality, region of residence, highest education level, field of study, employment status, employment sector, and organization size. These variables support subgroup analyses and help interpret differences in GenAI adoption and perceptions.

\paragraph{Awareness and Understanding.}
This domain assesses whether respondents have heard of GenAI tools (e.g., ChatGPT, DALL\textendash E), when they first became aware of them, and how they rate their understanding of what these tools do. Additional items probe familiarity with basic concepts, such as the idea that these tools generate content probabilistically, and ask whether respondents feel able to explain GenAI in their own words. These items provide the basis for distinguishing simple name recognition from deeper conceptual understanding.

\paragraph{Adoption and Usage Patterns.}
This domain examines whether and how respondents use GenAI tools. Questions cover user/non-user status, frequency of use (Never, Rarely, Monthly, Weekly, Daily), and typical purposes such as writing assistance, translation, coding support, creative generation, learning, and personal advice. Respondents also indicate the main contexts of use (personal, academic, or work). Non-users are asked to indicate reasons for non-adoption, such as lack of interest, lack of access, or low trust.

\paragraph{Perceived Impacts.}
This domain assesses how participants believe GenAI affects their tasks and skills. Using five-point scales ranging from \textit{Very Negative} to \textit{Very Positive}, respondents rate the impact of GenAI on speed of task completion, output quality, creativity, understanding complex information, skill acquisition, confidence in decision making, and critical thinking ability. A separate 0–10 scale asks how much reliance on GenAI negatively affects communication, writing, and problem-solving skills.

\paragraph{Concerns and Risk Perceptions.}
This domain explores trust in AI-generated content and perceived risks. Items ask how much respondents rely on GenAI outputs (from \textit{full trust without review} to \textit{no trust at all}) and invite them to select areas of concern, such as data privacy and security, content accuracy, ethical misuse, intellectual property, overreliance on AI, decline in personal skills, and job displacement. Additional items probe data-sharing practices, asking whether respondents or their peers have shared personal or sensitive information (e.g., names, email addresses, identification documents, financial data, health information) with GenAI tools.

\paragraph{Personal Reflections and Outlook.}
This domain captures broader attitudes and expectations regarding GenAI. One item asks participants to choose the statement that best describes their overall view of GenAI tools, ranging from strong optimism to strong skepticism. Three optional open-ended questions invite respondents to describe their personal impressions of using GenAI tools, to suggest initiatives or solutions that would benefit them or their communities, and to provide any additional comments.

\subsection{Data Collection Procedures}
We administered the survey online using Microsoft Forms between June and August 2025. Before beginning the survey, participants read an information statement explaining the study purpose, the voluntary nature of participation, and assurances of anonymity and confidentiality. Proceeding with the survey implied informed consent.

To ensure broad outreach, the survey link was distributed through multiple channels. Online, it was shared via WhatsApp, LinkedIn, and X (formerly Twitter). Offline, printed brochures with a QR code linking to the survey were circulated at public venues in Riyadh, including Boulevard City, the Sports Boulevard, and other high-traffic locations. This combined strategy helped reach respondents from different regions and backgrounds across Saudi Arabia.

\subsection{Data Preparation and Cleaning}
All responses were cleaned and harmonized before analysis. We first removed entries from non-Saudi nationals and incomplete submissions, yielding a final dataset of \(N = 330\) Saudi respondents. Arabic open-ended responses were translated into English to ensure a consistent working language.

Categorical variables with very sparse categories were merged into an ``Other'' category when appropriate, to avoid fragmentation. For multi-select questions, such as training topics, usage contexts, main uses, barriers, concerns, and types of data shared, we applied one-hot encoding so that each selected option is represented as a separate binary indicator (0 = not selected, 1 = selected). Prevalence for each option is computed as the column mean.

To support quantitative analysis, we mapped repeated Likert scales to numeric scores. Awareness and impact items share a five-point structure (e.g., \emph{Very low}, \emph{Low}, \emph{Average}, \emph{Elevated}, \emph{Very high} or \emph{Very negative} to \emph{Very positive}); these were mapped to values from 1 to 5. Usage frequency categories (Never, Rarely, Monthly, Weekly, Daily) were treated as ordered and recoded into an ordinal score from 0 to 4. Hours of GenAI use per week were cleaned, clipped to the range 0–80 to remove obvious outliers, and grouped into interpretable buckets (0, <1, 1–2, 3–5, 6–10, 11–20, 20+). 

We also derived simple status flags. Trust categories were mapped to a three-level numeric score (0 = no trust and full review, 2 = usually trust with verification, 3 = fully trust without review). For training and subscription status, responses of ``Yes'' or ``Sometimes'' were coded as 1, and ``No'' or unknown responses as 0, yielding binary indicators for wanting training and paying for GenAI tools. Missing numeric values were handled via pairwise deletion in statistical analyses, while missing categorical responses were retained and labeled as ``Unknown'' to preserve transparency in descriptive summaries.

\subsection{Quantitative Analysis}
We used descriptive and inferential statistics to address the quantitative components of the research questions.

\paragraph{Descriptive Statistics.}
We summarized all closed-ended items to describe GenAI adoption, use frequency, contexts of use, awareness, perceived impacts, training interests, and concerns. Categorical variables are reported using counts and percentages. Ordinal and numeric variables are described using means and standard deviations when distributions are approximately symmetric, and medians with interquartile ranges when they are skewed.

\paragraph{Composite Indices.}
To capture broader constructs, we constructed two composite scales. First, \textit{AwarenessMean} is defined as the average of three 5-point items measuring conceptual awareness of GenAI, technical understanding, and awareness of GenAI capabilities and limitations. Internal consistency was assessed using Cronbach’s \(\alpha\)~\cite{cronbach1951coefficient}, which showed excellent reliability (\(\alpha = 0.909\)). Second, the \textit{Perceived Impact Index} (PII) is defined as the mean of seven 5-point items measuring perceived effects of GenAI on speed of task completion, output quality, creativity, understanding complex information, skill improvement, decision confidence, and critical thinking. PII also demonstrated high reliability (\(\alpha = 0.893\)). For easier interpretation, we rescaled PII from its original 1–5 range to a 0–100 scale:
\[
\mathrm{PII}_{100} = \frac{(\mathrm{PII} - 1)}{4} \times 100.
\]

\paragraph{Bivariate Relationships.}
To explore how awareness and perceived impacts relate to background and usage factors, we used Spearman’s rank correlation (\(\rho\))~\cite{spearman1961proof} for ordinal and numeric predictors, such as use frequency, hours of use, trust scores, and outlook scores. For categorical predictors such as field of study, employment status, and sector, we used Kruskal–Wallis tests~\cite{kruskal1952use} to compare the distributions of \textit{AwarenessMean} and PII across groups. Alongside test statistics and \(p\)-values, we reported effect sizes (e.g., epsilon-squared) to indicate the magnitude of observed differences~\cite{tomczak2014need}.

\paragraph{Multivariate Modeling.}
To identify adjusted predictors of GenAI awareness, we fitted an ordinary least squares (OLS) regression model~\cite{montgomery2021introduction} with \textit{AwarenessMean} as the outcome. The model included demographic and usage-related covariates, with categorical variables entered as dummy indicators. Because survey data often exhibit heteroskedasticity, we used HC3 robust standard errors~\cite{mackinnon1985some}. This approach allowed us to assess which factors remained significantly associated with awareness after accounting for other variables.

\paragraph{Risk Perceptions, Barriers, and Data Sharing.}
To analyze group differences in perceived risks and barriers (e.g., privacy concerns, worries about misinformation, job loss, or overreliance), we used one-way ANOVA~\cite{fisher1970statistical} to compare mean concern levels across age groups, fields of study, employment statuses, and sectors. For each comparison, we reported the \(F\)-statistic, \(p\)-value, and eta-squared (\(\eta^2\)) as a measure of variance explained. For data-sharing practices, we computed mean probabilities of sharing different data types (e.g., email address, date of birth, health data) with GenAI tools for each employment group and visualized these using bar plots. While these comparisons are mainly descriptive, they provide insight into employment-based differences in disclosure behavior.

\paragraph{Training Topics and Co-Selection Patterns.}
To address training interests, we examined multi-select responses on preferred GenAI training topics and tools. Using the one-hot encoded variables, we computed pairwise Jaccard similarity~\cite{travieso2023jaccard} to quantify how often two topics were selected together relative to how often either was selected. This analysis highlights natural bundles of training needs, such as combinations of privacy and ethical issues, or specific AI coding assistants, which inform our discussion of curriculum and policy implications.

\subsection{Qualitative Analysis}
Open-ended responses, mainly associated with personal impressions, desired initiatives, and final reflections, were analyzed using inductive thematic analysis~\cite{braun2006using}. After translation and cleaning, all responses were read multiple times to achieve familiarity with the data. We then generated initial codes that captured recurring ideas, including perceived benefits, challenges, cultural and ethical concerns, calls for Arabic-first tools, expectations for regulation, and views on equity and access.

Codes were iteratively grouped into broader themes corresponding to the survey domains. We examined how often each theme appeared and selected illustrative quotations to represent participants’ perspectives. These qualitative findings are used in the Results section to enrich and contextualize the quantitative trends, providing a more complete picture of how GenAI is perceived, adopted, and trusted in Saudi Arabia.
 
\section{Results}
\label{sec:Results}

This section presents key findings from our national survey (\(N = 330\)), progressing from descriptive insights to deeper explanatory analyses. We begin by outlining the demographic composition of participants, followed by their levels of adoption, usage frequency, and training interests related to GenAI. Next, we present results on awareness and understanding, including trust levels, perceptions of content accuracy, and engagement with safeguards. To quantify users’ experiences, we introduce a composite Perceived Impact Index (PII), derived from seven Likert-scale items, which captures the broader societal and personal effects of GenAI use.

The subsequent sections examine differences across demographic and occupational groups. Using one-way ANOVA, we identify statistically significant variations in perceived barriers (e.g., trust, usability), concerns (e.g., job loss, misinformation), and willingness to share personal data (e.g., email, date of birth), offering contextual insights into how adoption risks and behaviors are shaped by age, education, employment status, and sector. Finally, we analyze open-ended responses using thematic coding to uncover user sentiments, expectations, and desired future directions for GenAI in Saudi Arabia.

\subsection{RQ1 – Adoption Frequency, Contexts, and Use Cases}

\begin{figure}[t]
\centering
\begin{tikzpicture}
\begin{axis}[
  ybar,
  height=6.0cm,
  width=0.9\linewidth,
  ymin=0, ymax=50,
  ylabel={Percent of respondents},
  symbolic x coords={Before 2020,2020--2021,2022--2023,2024--2025,Never used},
  xtick=data,
  enlarge x limits=0.05,
  nodes near coords,
  nodes near coords align={vertical},
  bar width=14pt,
  xticklabels={
  Before 2020\\{\scriptsize (n=21)},
  2020--2021\\{\scriptsize (n=45)},
  2022--2023\\{\scriptsize (n=126)},
  2024--2025\\{\scriptsize (n=114)},
  Never Used\\{\scriptsize (n=24}
},
  x tick label style={
    font=\footnotesize,
    rotate=15,
    anchor=north east,
    xshift=18pt,
    align=center, 
  }
]
\addplot coordinates {
  (Before 2020,6.36)
  (2020--2021,13.64)
  (2022--2023,38.18)
  (2024--2025,34.55)
  (Never used,7.27)
};
\end{axis}
\end{tikzpicture}
\caption{Timing of first GenAI use among all respondent. (\(N=330\))}
\label{fig:first_use_timing}
\end{figure}
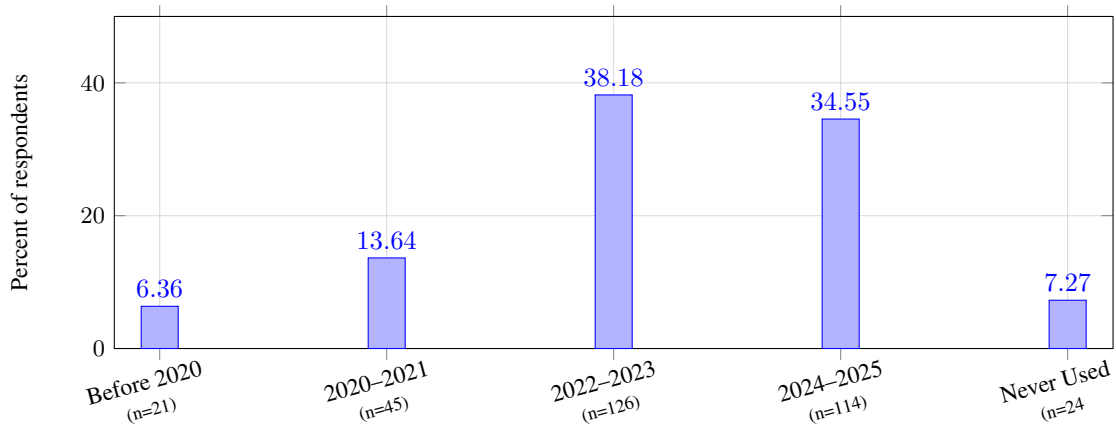

\begin{figure}[t]
\centering
\begin{tikzpicture}
\begin{axis}[
  ybar, height=6.0cm, width=0.9\linewidth,
  ymin=0, ymax=50,
  ylabel={Percent of respondents},
  symbolic x coords={Rarely,Monthly,Weekly,Daily},
  xtick=data, enlarge x limits=0.05,
  nodes near coords, nodes near coords align={vertical},
  bar width=14pt,
  xticklabels={
  Rarely\\{\scriptsize (n=33)},
  Monthly\\{\scriptsize (n=19)},
  Weekly\\{\scriptsize (n=115)},
  Daily\\{\scriptsize (n=139)}
},
  x tick label style={
    font=\footnotesize,
    rotate=15,
    anchor=north east,
    xshift=13pt,
    align=center, 
  }
]

\addplot coordinates {
  (Daily,10.78)
  (Weekly,6.21)
  (Monthly,37.58)
  (Rarely,45.43)
};

\end{axis}
\end{tikzpicture}
\caption{Distribution of use frequency among GenAI users. (\(N=306\))}
\label{fig:usefreq_dist}
\end{figure}
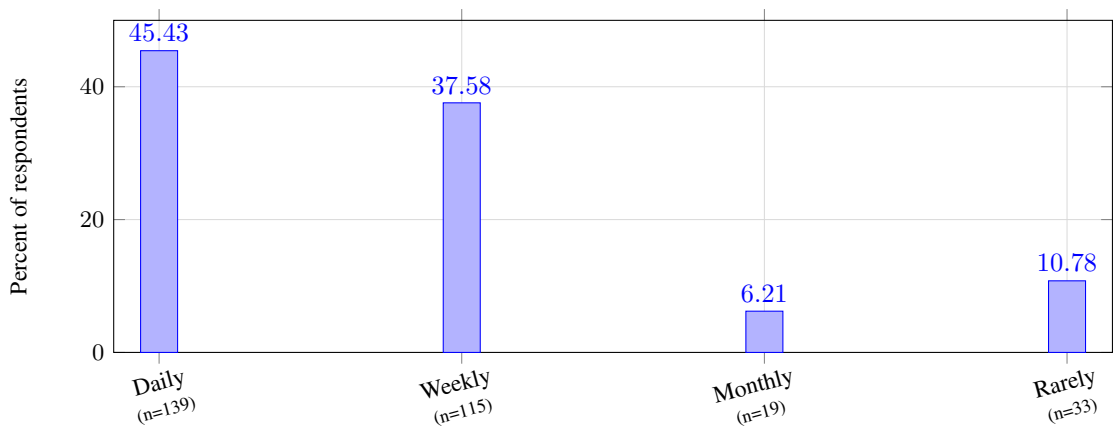


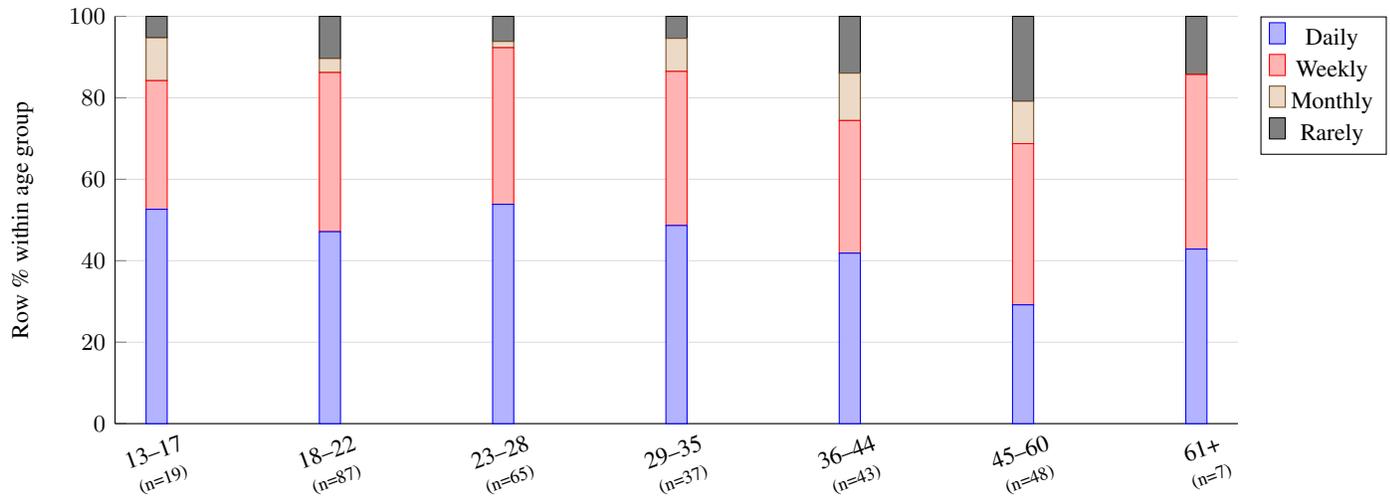
\begin{figure}[t]
\centering
\begin{tikzpicture}
\begin{axis}[
  ybar stacked,
  height=7.0cm,
  width=\linewidth,
  ymin=0, ymax=100,
  ylabel={Row \% within age group},
  symbolic x coords={13–17,18–22,23–28,29–35,36–44,45–60,61+},
  xtick=data,
  xticklabels={
  13--17\\{\scriptsize (n=19)},
  18--22\\{\scriptsize (n=87)},
  23--28\\{\scriptsize (n=65)},
  29--35\\{\scriptsize (n=37)},
  36--44\\{\scriptsize (n=43)},
  45--60\\{\scriptsize (n=48)},
  61+\\{\scriptsize (n=7)}
},
  x tick label style={ 
  font=\footnotesize,
  rotate=20,
  anchor=north east,
  yshift=-3pt,
  xshift=10pt,
  align=center, 
},
  axis x line*=bottom,
  axis y line*=left,
  enlarge x limits=0.04,
  bar width=8pt,
  legend style={
    at={(1.02,1)},
    anchor=north west,
    font=\footnotesize
  }
]
\addplot coordinates {
  (13–17,52.63) (18–22,47.13) (23–28,53.85) (29–35,48.65)
  (36–44,41.86) (45–60,29.17) (61+,42.86)
};
\addlegendentry{Daily}

\addplot coordinates {
  (13–17,31.58) (18–22,39.08) (23–28,38.46) (29–35,37.84)
  (36–44,32.56) (45–60,39.58) (61+,42.86)
};
\addlegendentry{Weekly}

\addplot coordinates {
  (13–17,10.53) (18–22,3.45) (23–28,1.54) (29–35,8.11)
  (36–44,11.63) (45–60,10.42) (61+,0.00)
};
\addlegendentry{Monthly}

\addplot coordinates {
  (13–17,5.26) (18–22,10.34) (23–28,6.15) (29–35,5.41)
  (36–44,13.95) (45–60,20.83) (61+,14.29)
};
\addlegendentry{Rarely}
\end{axis}
\end{tikzpicture}
\caption{Use frequency by age group (row percentages). (\(N=306\))}
\label{fig:usefreq_by_age}
\end{figure}

  The adoption of GenAI tools in Saudi Arabia is both widespread and increasingly embedded in everyday digital habits. Figure~\ref{fig:first_use_timing} shows that most respondents began using GenAI very recently, with 72.7\% reporting their first use between 2022 and 2025. This pattern aligns with the global surge of interest following the public release of ChatGPT in late 2022, suggesting that mainstream availability and visibility of advanced GenAI tools played a major role in driving adoption. A smaller share (20\%) experimented with GenAI before 2022, while 24 respondents (7.27\%) reported never having used any GenAI tool.
  
 As shown in Figure~\ref{fig:usefreq_dist},  about 83\% of GenAI users report using GenAI at least weekly, with 45\% indicating daily use and 38\% reporting weekly use. Only around 11\% report using GenAI ‘Rarely’ and 6\% ‘Monthly’. These results indicate that GenAI is no longer a niche technology but a routine part of many users’ digital lives.

Figure~\ref{fig:usefreq_by_age} shows how usage frequency varies across age groups among GenAI users (\(N=306\)). Across all groups, most respondents report using GenAI at least weekly, with the combined share of daily and weekly users ranging from about 69\% among adults aged 45--60 to over 90\% among those aged 23--28. Young adults (18--28) stand out as particularly intensive users, with roughly half reporting daily use and another 38--39\% using GenAI weekly. Older adults display more varied engagement patterns: for example, in the 45--60 group, around 29\% report daily use, 40\% weekly use, and about one third use GenAI only monthly or rarely. The oldest group (61+), although small in size (\(n=7\)), also shows high engagement, with nearly all respondents using GenAI at least weekly (split roughly evenly between daily and weekly use). These patterns suggest that frequent GenAI use is not confined to younger participants, even though the highest concentrations of intensive users are found among young adults.

\begin{figure}[t]
\centering
\begin{tikzpicture}
\begin{axis}[
  ybar stacked,
  height=7.0cm,
  width=\linewidth,
  ymin=0, ymax=100,
  ylabel={Row \% within age group},
  symbolic x coords={13–17,18–22,23–28,29–35,36–44,45–60,61+},
  xtick=data,
  xticklabels={
  13--17\\{\scriptsize (n=16)},
  18--22\\{\scriptsize (n=75)},
  23--28\\{\scriptsize (n=60)},
  29--35\\{\scriptsize (n=32)},
  36--44\\{\scriptsize (n=32)},
  45--60\\{\scriptsize (n=33)},
  61+\\{\scriptsize (n=6)}
},
  x tick label style={ 
  font=\footnotesize,
  rotate=20,
  anchor=north east,
  yshift=-3pt,
  xshift=10pt,
  align=center, 
},
  axis x line*=bottom,
  axis y line*=left,
  enlarge x limits=0.04,
  bar width=8pt,
  legend columns=1,
  legend cell align=left,
  legend style={
    at={(1.02,1)},
    anchor=north west,
    font=\footnotesize
  }
]

\addplot coordinates {
  (13–17,43.75) (18–22,22.67) (23–28,35.00) (29–35,21.88)
  (36–44,31.25) (45–60,42.42) (61+,16.67)
};
\addlegendentry{\textless{}1 h}

\addplot coordinates {
  (13–17,43.75) (18–22,50.67) (23–28,45.00) (29–35,37.50)
  (36–44,50.00) (45–60,39.39) (61+,50.00)
};
\addlegendentry{1--5 h}

\addplot coordinates {
  (13–17,6.25) (18–22,14.67) (23–28,13.33) (29–35,18.75)
  (36–44,12.50) (45–60,12.12) (61+,16.67)
};
\addlegendentry{6--10 h}

\addplot coordinates {
  (13–17,6.25) (18–22,6.67) (23–28,3.33) (29–35,6.25)
  (36–44,6.25) (45–60,3.03) (61+,0.00)
};
\addlegendentry{11--20 h}

\addplot coordinates {
  (13–17,0.00) (18–22,5.33) (23–28,3.33) (29–35,15.62)
  (36–44,0.00) (45–60,3.03) (61+,16.67)
};
\addlegendentry{\textgreater{}20 h}

\end{axis}
\end{tikzpicture}
\caption{Self-reported GenAI usage hours per week by age group (row percentages). Across respondents who reported weekly hours. (\(N=254\))}
\label{fig:hours_by_age}
\end{figure}
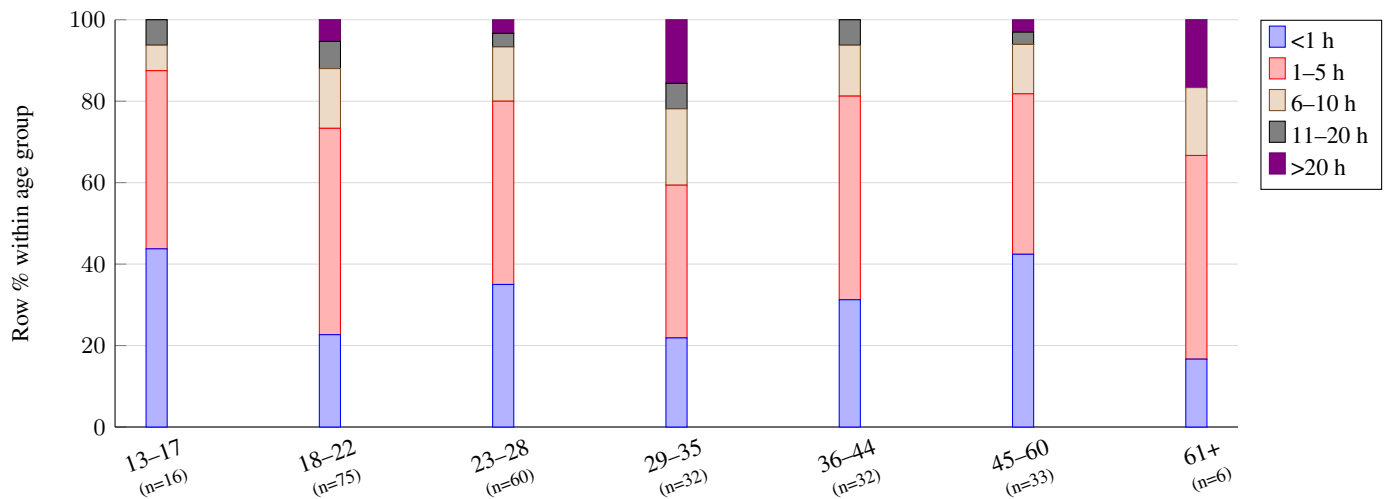

To better understand engagement levels, respondents indicated how many hours per week they typically spend using GenAI tools by selecting one of five ranges: \textless{}1 hour, 1--5 hours, 6--10 hours, 11--20 hours, or more than 20 hours per week (Figure~\ref{fig:hours_by_age}). Across the full sample (\(N=254\)), usage tends to be modest: around three quarters of participants reported using GenAI for 5 hours per week or less, with the 1--5 hour range being the most common category in every age group. University-aged respondents (18--22) showed somewhat higher engagement, with about half reporting 1--5 hours per week and roughly one quarter reporting 6 or more hours. Adults aged 29--35 contained the largest share of heavier users (6+ hours per week), including around 16\% who reported more than 20 hours. The oldest group (61+), although small in size (\(n=6\)), also included both low- and high-intensity users, with half reporting 1--5 hours per week and about one third reporting 6 or more hours. 

Overall, these patterns suggest that GenAI is used regularly but not intensively by most respondents, and that moderate-to-high engagement is present across the adult life span rather than being confined to the youngest participants.

\begin{table}[t]
\centering
\caption{Primary contexts in which respondents report using GenAI tools (multiple responses allowed). (\(N = 306\))}
\label{tab:use_context}
\begin{tabular}{lrr}
\toprule
Context                       & Count (n) & Percent of users \\
\midrule
Personal use                  & 246       & 80.4\%           \\
Learning / academic study     & 194       & 63.4\%           \\
Work / job-related tasks      & 152       & 49.7\%           \\
Other & 1         & 0.3\%            \\
\bottomrule
\end{tabular}
\end{table}

To characterize where GenAI fits into respondents' everyday lives, we also asked in which contexts they typically use GenAI, allowing multiple selections among personal use, learning or academic study, work or job-related tasks, and an open-ended "Other" option. As shown in Table~\ref{tab:use_context}, personal use is the dominant context, reported by 80.4\% of GenAI users, followed by learning or academic study (63.4\%) and work- or job-related tasks (49.7\%). Because these categories are not mutually exclusive, many respondents report using GenAI across several spheres of life, combining personal experimentation, study support, and work-related productivity. 

This pattern suggests that, at least in our sample, GenAI has diffused first into personal and self-directed learning activities rather than being confined to formal workplace tools, even as nearly half of users already integrate GenAI into their jobs. Only one respondent indicated using GenAI exclusively for “other” purposes, underscoring that most use is anchored in these three main domains.

\begin{figure}[htbp]
\centering
\begin{tikzpicture}
\begin{axis}[
    xbar,
    width=0.9\textwidth,
    height=8.5cm,
    bar width=6pt,
    enlarge y limits=0.05,
    xlabel={Percentage (\%)},
    symbolic y coords={
        {Audio Generation},
        {Video Generation},
        {Creative Writing},
        {Image Generation},
        {Programming Assistance},
        {Translation},
        {Data Analysis},
        {Learning New Concepts},
        {Information Summarization},
        {Brainstorming},
        {Content Writing},
        {Research Assistance}
    },
    ytick=data,
    nodes near coords,
    nodes near coords align={horizontal},
    every node near coord/.append style={font=\footnotesize},
    xtick={0,10,...,70},
    xmin=0, xmax=70,
    tick style={draw=none},
    xmajorgrids=true,
    ytick style={draw=none},
    axis line style={draw=none},
    tick label style={font=\footnotesize},
    ylabel={}
]
\addplot+[xbar, fill=myblue] coordinates {
    (7.6,{Audio Generation})
    (9.7,{Video Generation})
    (22.7,{Creative Writing})
    (23.6,{Image Generation})
    (29.4,{Programming Assistance})
    (31.5,{Translation})
    (43.9,{Data Analysis})
    (44.6,{Learning New Concepts})
    (50.9,{Information Summarization})
    (52.7,{Brainstorming})
    (54.9,{Content Writing})
    (62.4,{Research Assistance})
};
\end{axis}
\end{tikzpicture}
\caption{Common usage scenarios of GenAI across participants. (\(N = 306\))}
\label{fig:genai_tasks}
\end{figure}
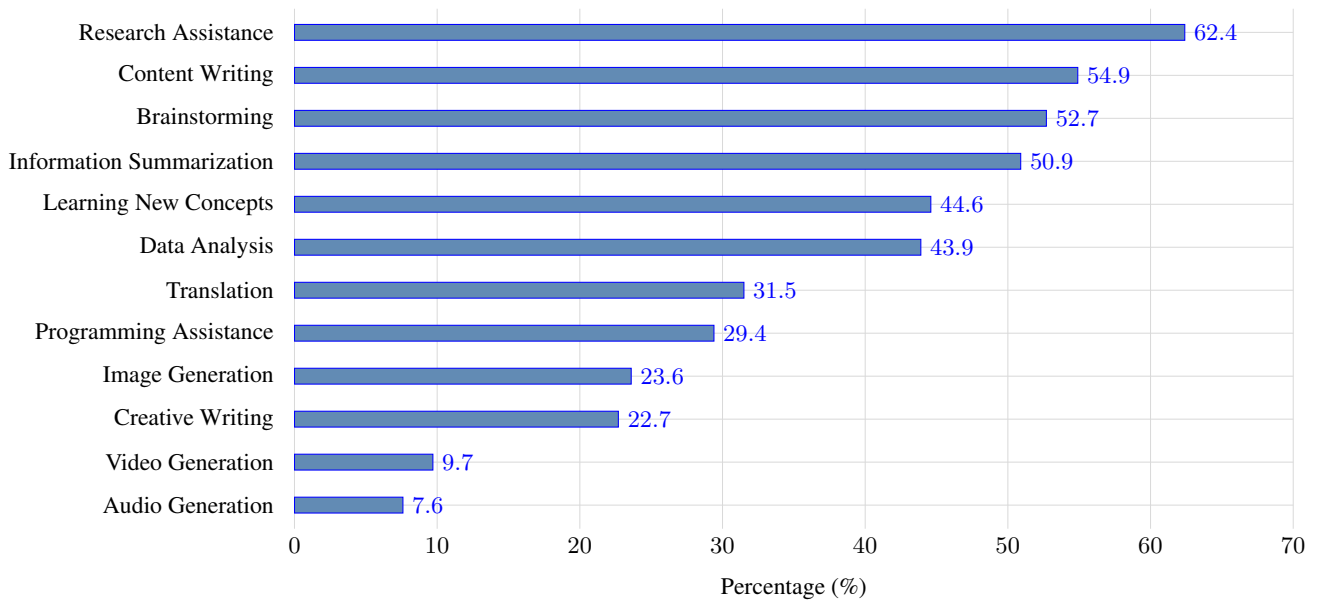

We also examine how respondents use GenAI in their daily lives at the level of specific activities. Within the personal, academic, and work contexts described above, the dominant use cases are text- and knowledge-focused (Figure~\ref{fig:genai_tasks}). The most common activities included assistance with research (62.4\%), content writing such as emails, reports, and similar documents (54.9\%), and brainstorming or idea generation (52.7\%). Overall, these results indicate that current GenAI usage in Saudi Arabia is heavily centered on text-based productivity, learning support, and knowledge work, rather than on multimedia content creation.

\begin{tcolorbox}[colback=blue!5!white, colframe=blue!75!black,
title=Summary: GenAI Adoption, fonttitle=\bfseries]
GenAI use in Saudi Arabia is widespread and has accelerated rapidly in recent years. Most respondents began using GenAI after 2022, following the public release of tools like ChatGPT, while 7\% have not yet used GenAI. Among actual users (N=306), approximately 83\% engage with GenAI at least weekly, including 45\% who use it daily. Young adults (18–28) remain the most intensive users, whereas older adults show more varied engagement patterns. Across age groups, most respondents report relatively light weekly use typically 5 hours or less with a smaller subset of heavier users who report 6 or more hours per week. Usage spans multiple contexts, led by personal use (80\%), academic learning (63\%), and work-related tasks (50\%), and  primarily used for research support, writing, summarization, and brainstorming, with more limited adoption of multimedia generation tasks such as images, video, or sound.
\end{tcolorbox}

\subsection{RQ2 – Awareness and Technical Understanding of GenAI}
\label{sec:Awareness}

\begin{figure}[ht]
\centering
\begin{tikzpicture}
\begin{axis}[
    ybar,
    bar width=10pt,
    width=\textwidth,
    height=8cm,
    ymin=0, ymax=40,
    enlarge x limits=0.2,
    symbolic x coords={Very Low,Low,Average,Elevated,Very High},
    xtick=data,
    xlabel={Awareness/Understanding Level},
    ylabel={\% of Participants},
    legend style={at={(0.5,-0.15)}, anchor=north, legend columns=3},
    nodes near coords,
    nodes near coords align={vertical},
    every node near coord/.append style={font=\tiny}
]
\addplot+[fill=gray!30] coordinates {
    (Very Low,11.52)
    (Low,15.45)
    (Average,37.88)
    (Elevated,22.12)
    (Very High,13.03)
};
\addlegendentry{Concept Awareness}

\addplot+[fill=gray!60] coordinates {
    (Very Low,14.55)
    (Low,24.55)
    (Average,36.06)
    (Elevated,14.55)
    (Very High,10.30)
};
\addlegendentry{Technical Understanding}

\addplot+[fill=gray!80] coordinates {
    (Very Low,10.91)
    (Low,16.06)
    (Average,33.03)
    (Elevated,24.55)
    (Very High,15.45)
};
\addlegendentry{Capabilities \& Limits Awareness}

\end{axis}
\end{tikzpicture}
\caption{Distribution of responses across five levels of awareness and understanding of GenAI. (\(N=330\))}
\label{fig:awareness_levels}
\end{figure}
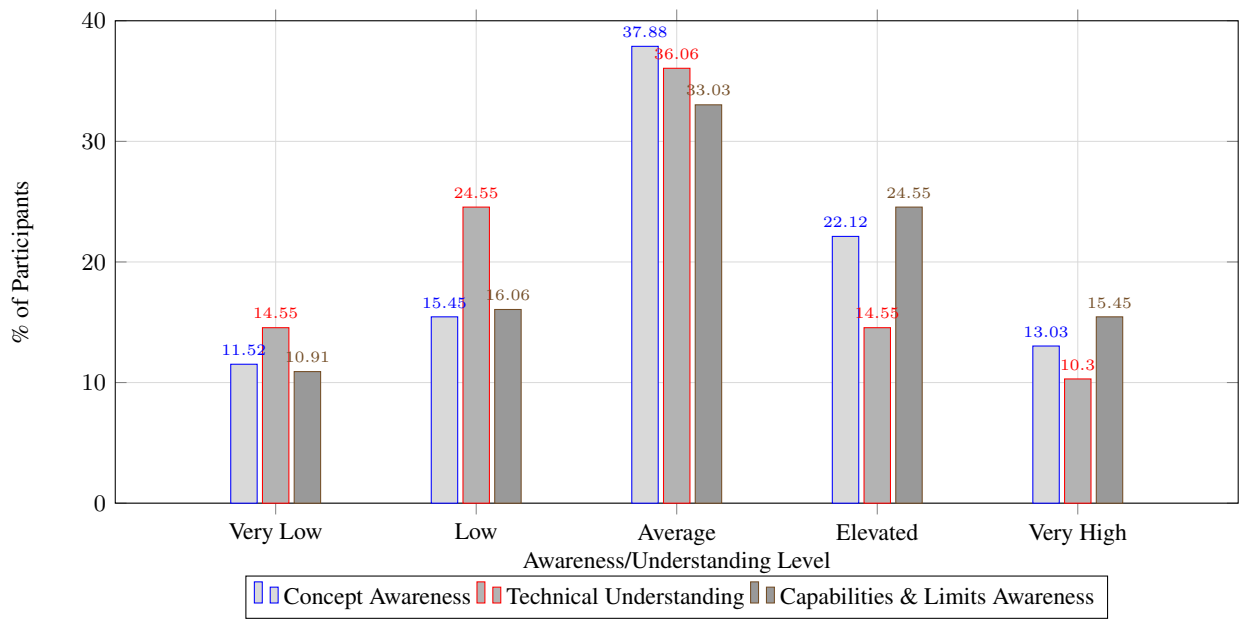

To evaluate how familiar the public in Saudi Arabia is with GenAI, we focused on three related dimensions: 
(1) \emph{conceptual awareness}, a general sense of what GenAI is and how it works; 
(2) \emph{technical understanding}, knowledge of the underlying mechanisms that power GenAI; and 
(3) \emph{awareness of capabilities and limitations}, an understanding of what GenAI can and cannot do.

Each dimension was measured on a five-point scale from 1 (Very Low) to 5 (Very High). To confirm that these three items formed a reliable scale, we assessed internal consistency using Cronbach’s alpha~\cite{cronbach1951coefficient}. The result (\(\alpha = 0.909\)) indicated excellent reliability, allowing us to combine them into a single awareness index.

As shown in Figure~\ref{fig:awareness_levels}, most participants reported moderate to high levels of conceptual awareness. Specifically, 37.9\% rated their awareness as ``Average,'' 22.1\% as ``Elevated,'' and 13.0\% as ``Very High.'' Only 11.5\% rated their awareness as ``Very Low.'' This suggests that most respondents have at least a basic understanding of what GenAI is, reflecting its growing presence in public discussions.

However, when it came to technical understanding, the results were more modest. While 36.1\% chose ``Average,'' a large portion reported ``Low'' (24.6\%) or ``Very Low'' (14.6\%) understanding. Fewer than 11\% rated their technical grasp as ``Very High.'' This gap suggests that while many are aware of GenAI in general terms, fewer people understand how it works under the hood.

Responses to the third dimension, awareness of GenAI's capabilities and limitations were slightly more optimistic. About one-third (33.0\%) rated their awareness as ``Average,'' 24.6\% as ``Elevated,'' and 15.5\% as ``Very High.'' This shows that even if people lack deep technical knowledge, many still have a practical sense of what GenAI can and cannot do, indicating a balanced understanding that includes both potential and limits.

Descriptive statistics for these three items are presented in Table~\ref{tab:awareness_stats}. On average, participants reported the highest awareness for GenAI’s capabilities and limitations (mean = 3.18), followed by conceptual awareness (mean = 3.10). Technical understanding received the lowest average score (mean = 2.82), reinforcing the idea that surface-level familiarity is more common than deeper comprehension.

\begin{table}[h]
\centering
\caption{Descriptive statistics for awareness and understanding items (\(N=330\))}
\label{tab:awareness_stats}
\begin{tabular}{@{}lcccccccc@{}}
\toprule
\textbf{Item} & \textbf{Count} & \textbf{Mean} & \textbf{SD} & \textbf{Min} & \textbf{25\%} & \textbf{Median} & \textbf{75\%} & \textbf{Max} \\
\midrule
Concept Awareness                & 330 & 3.10 & 1.16 & 1.00 & 2.00 & 3.00 & 4.00 & 5.00 \\
Technical Understanding         & 330 & 2.82 & 1.16 & 1.00 & 2.00 & 3.00 & 3.00 & 5.00 \\
Capabilities \& Limits Awareness & 330 & 3.18 & 1.20 & 1.00 & 2.00 & 3.00 & 4.00 & 5.00 \\
\bottomrule
\end{tabular}
\end{table}

\begin{table}[h]
\centering
\caption{Drivers of Awareness (AwarenessMean as outcome). Bivariate tests use Spearman's $\rho$ for numeric/ordinal and Kruskal–Wallis (KW) for categorical predictors. Adjusted model: OLS with HC3 robust SEs.}
\label{tab:drivers_awareness}
\scriptsize
\begin{tabular}{@{}lccccccc@{}}
\toprule
\textbf{Predictor} & \textbf{Bivariate Test} & \textbf{Effect / Stat} & \textbf{Biv. p-value} & \textbf{Effect Size ($\varepsilon^2$ or ---)} & \textbf{Adj. Stat} & \textbf{Direction (Biv.)} & \textbf{Adj. p-value} \\
\midrule
Use frequency                & Spearman $\rho$      & 0.406   & <0.001 & ---   & 0.299   & ↑ with Awareness      & <0.001 \\
Trust level                 & Spearman $\rho$      & 0.032   & 0.566  & ---   & 0.038   & ↑ with Awareness      & 0.506  \\
Barriers               & Spearman $\rho$      & 0.012   & 0.822  & ---   & -0.045  & ↑ with Awareness      & 0.494  \\
Concerns               & Spearman $\rho$      & 0.061   & 0.271  & ---   & 0.019   & ↑ with Awareness      & 0.566  \\
First experience (groups)   & KW H (k=2)           & 22.029  & <0.001 & 0.064 & 2.350   & varies by group       & 0.125  \\
Education level             & KW H (k=5)           & 4.054   & 0.399  & 0.000 & 0.694   & varies by group       & 0.952  \\
Field of study              & KW H (k=10)          & 40.994  & <0.001 & 0.100 & 15.273  & varies by group       & 0.084  \\
Employment status           & KW H (k=5)           & 18.841  & 0.001  & 0.046 & 4.877   & varies by group       & 0.300  \\
Sector (Top-4 + Other)      & KW H (k=4)           & 21.887  & <0.001 & 0.058 & 8.914   & varies by group       & 0.030  \\
\bottomrule
\end{tabular}
\end{table}

To simplify interpretation, we created a single metric, \textit{AwarenessMean} by averaging each respondent’s scores across the three dimensions. The resulting distribution centered around the midpoint of the scale (\(M = 3.03\), \(SD = 1.08\); median = 3.00; IQR = 2.33–3.67). We then grouped respondents into three categories: Low awareness (<3; 58.8\%), Medium awareness (3 to <4; 27.9\%), and High awareness (\(\geq 4\); 13.3\%).

\vspace{0.5em}
To better understand what drives GenAI awareness, we conducted a two-stage analysis:

\begin{itemize}
  \item In the first stage, we used bivariate statistical tests. For ordinal and numeric variables (e.g., GenAI usage frequency, trust, and reported barriers), we applied Spearman’s rank correlation (\(\rho\))~\cite{spearman1961proof}. For categorical variables (e.g., education, employment sector), we used Kruskal–Wallis (KW) tests~\cite{kruskal1952use}, reporting effect sizes using epsilon-squared (\(\varepsilon^2\))~\cite{tomczak2014need}.
  \item In the second stage, we built an adjusted model using ordinary least squares (OLS) regression with HC3 robust standard errors~\cite{montgomery2021introduction, mackinnon1985some}. The outcome variable was \textit{AwarenessMean}. Numeric predictors were treated as continuous, while categorical ones were modeled with indicator variables and tested via Wald \(\chi^2\) statistics. To reduce noise and improve interpretability, we converted multi-select responses into count variables (e.g., number of concerns) and grouped employment sectors into a ``Top-4 + Other'' format.
\end{itemize}

\vspace{0.5em}
Table~\ref{tab:drivers_awareness} revealed that GenAI awareness grows through \emph{use}, not just exposure. Participants who used GenAI more frequently showed significantly higher awareness in both bivariate and adjusted models. In the bivariate test, use frequency had a strong positive correlation with awareness (\(\rho = 0.406\), \(p < .001\)). In the adjusted regression, usage remained the strongest predictor (\(\beta = 0.299\), \(p < .001\)). This means that each increase in use frequency (e.g., from ``Rarely'' to ``Sometimes'') was linked to an average 0.30-point increase in awareness. These results suggest that hands-on experience, not just hearing about GenAI, is what deepens understanding.

Employment sector was the only contextual factor that remained significant in the adjusted model (Wald \(\chi^2 = 8.91\), \(p = .030\)). Although other factors, such as field of study (\(\varepsilon^2 \approx 0.10\)) and first point of exposure (\(\varepsilon^2 \approx 0.06\)), showed significance in the bivariate stage, their effects did not hold after adjusting for other variables. This implies that certain professional environments may provide better exposure to GenAI, reinforcing awareness through daily tasks and informal learning.

In contrast, variables like trust in GenAI, formal education, barriers, and concerns had no significant impact in the adjusted model. This suggests that GenAI awareness is shaped more by practical experience and workplace context than by academic background or general attitudes.

\begin{tcolorbox}[colback=blue!5!white, colframe=blue!75!black,
title=Summary: GenAI Awareness, fonttitle=\bfseries]

Overall GenAI awareness among Saudi participants was moderate. Awareness was highest for understanding capabilities and lowest for technical knowledge. About 59\% of respondents had low overall awareness, while only 13\% scored high. The most important factor influencing awareness was frequent GenAI use, followed by employment sector. In contrast, education level, trust, and concerns did not show significant effects, suggesting that experience and context matter more than background or beliefs.

\end{tcolorbox}

\subsection{RQ3 – Perceived Impacts of GenAI}
To understand how users perceive the benefits of GenAI, we measured seven key dimensions of impact: 
\emph{speed of task completion}, 
\emph{output quality}, 
\emph{creativity of ideas}, 
\emph{understanding complex information}, 
\emph{skill development}, 
\emph{decision confidence}, and 
\emph{critical thinking}. 

Each dimension was rated on a 5-point Likert scale, from 1 (Very Low) to 5 (Very High). As with the awareness items in Section~\ref{sec:Awareness}, we tested internal consistency across these seven items. The results showed excellent reliability (Cronbach’s \(\alpha = 0.893\)), supporting their aggregation into a single summary measure.

\vspace{0.5em}
We constructed a composite metric called the \emph{Perceived Impacts Index (PII)} by averaging each respondent’s ratings across the seven items, requiring at least five non-missing responses. To make the results easier to interpret, we also converted PII into a 0–100 scale using the following formula:

\[
\text{PII}_{100} = \frac{(\text{PII} - 1)}{4} \times 100.
\]

Table~\ref{tab:impact_metrics} provides detailed statistics for each impact dimension and for the overall PII. For each, we report the sample size, mean and median ratings, the percentage of respondents rating the item \(\geq 4\) (indicating high or very high perceived benefit), and the normalized mean on the 0–100 scale.

\vspace{0.5em}
Participants generally viewed GenAI as having a positive impact, especially on efficiency and comprehension. Speeding up task completion stands out as the top benefit, it has the highest mean rating (4.04), the highest percentage of high/very high ratings (71.6\%), and the top normalized score (76.1/100). Understanding complex information is the second-strongest dimension (mean = 3.83, 64.4\% $\geq$ 4, score = 70.7), suggesting that many users see GenAI as a support tool for learning, exploring new topics, or making sense of difficult material. 

Skill improvement (mean = 3.76) and creativity of ideas (mean = 3.73) are also areas of notable benefit, with about 61\% of respondents in each case rating the impact as high or very high and normalized scores around the high 60s, indicating that users frequently credit GenAI with helping them develop abilities and generate new ideas.

\definecolor{low}{RGB}{255,245,235}     
\definecolor{mid}{RGB}{255,200,150}     
\definecolor{high}{RGB}{255,100,0}      
\begin{table}[ht]
\centering
\caption{Perceived Impact of GenAI on User Capabilities}
\label{tab:impact_metrics}
\scriptsize
\begin{tabular}{@{}lrrrr>{\columncolor{low!0!high}}r@{}}
\toprule
\textbf{Dimensions} & \textbf{n} & \textbf{Mean (1–5)} & \textbf{Median} & \textbf{\% $\geq$ 4} & \textbf{Alt. Mean (0–100)} \\
\midrule
Impact — Speed of Task Completion         & 306 & 4.042 & 4.000 & 71.6\% & \cellcolor{high!80}76.1 \\
Impact — Understanding Complex Information & 306 & 3.827 & 4.000 & 64.4\% & \cellcolor{high!70}70.7 \\
Impact — Skill Improvement                & 306 & 3.758 & 4.000 & 60.8\% & \cellcolor{high!65}69.0 \\
Impact — Creativity of Ideas              & 306 & 3.729 & 4.000 & 60.8\% & \cellcolor{high!64}68.2 \\
Impact — Output Quality                   & 306 & 3.621 & 4.000 & 53.3\% & \cellcolor{high!58}65.5 \\
Impact — Critical Thinking                & 306 & 3.431 & 3.000 & 44.8\% & \cellcolor{high!44}60.8 \\
Impact — Decision Confidence              & 306 & 3.425 & 3.000 & 44.1\% & \cellcolor{high!42}60.6 \\
\midrule
PII (composite, 1–5)                      & 306 & 3.690 & 3.710 & 57.1\% & \cellcolor{high!60}67.3 \\
\bottomrule
\end{tabular}
\end{table}

Other capabilities show more modest but still positive perceived gains. Output quality is viewed favorably (mean = 3.62, median = 4.00), with 53.3\% rating this dimension $\geq$ 4 and a normalized score of 65.5, implying that GenAI often helps users refine or polish their work but not as consistently as it accelerates tasks or aids comprehension.

However, the lowest ratings are found for critical thinking (mean = 3.43, 44.8\% $\geq$ 4, score = 60.8) and decision confidence (mean = 3.43, 44.1\% $\geq$ 4, score = 60.6). Fewer than half of respondents perceived strong benefits in these reasoning-oriented areas, and their medians drop to 3. This pattern suggests that while GenAI is widely seen as a powerful tool for speeding up work, understanding information, and enhancing skills and creativity, users are more cautious about its contribution to higher-order judgment, independent reasoning, and confidence in their final decisions.

The overall composite PII score averaged 3.69 (median = 3.71), which translates to 67.3 on the 0–100 scale. Around 57.1\% of participants had a PII score of 4 or higher, reflecting a generally favorable but not universal perception of GenAI's impact.

\begin{tcolorbox}[colback=blue!5!white, colframe=blue!75!black,
title=Summary: Perceived Impacts of GenAI, fonttitle=\bfseries]

Participants reported strong perceived gains in task speed and information comprehension, with additional benefits in creativity and skill development. However, the perceived impact was weaker in areas that require deeper reflection and independent thinking, such as critical reasoning and decision-making. To maximize benefits, future interventions should preserve GenAI's strengths in productivity while offering targeted support for cognitive and analytical skills.

\end{tcolorbox}

\subsection{RQ4 – Training Interests and Co-selected GenAI Topics}
To understand participants’ priorities for future GenAI training, we asked them to select topics of interest from a predefined list. Table~\ref{tab:training_overview} presents the most frequently selected training topics.

\begin{figure}[!htbp]
\centering
\begin{tikzpicture}
\begin{axis}[
  ybar, height=6.0cm, width=0.9\linewidth,
  ymin=0, ymax=70,
  ylabel={Percent of respondents},
  symbolic x coords={Definite,Possible,No/Other},
  xtick=data, enlarge x limits=0.10,
  nodes near coords, nodes near coords align={vertical},
  bar width=14pt,
  xticklabels={
    Definite\\{\scriptsize (n=190)},
    Possible\\{\scriptsize (n=77)},
    No/Other\\{\scriptsize (n=63)}
  },
  x tick label style={
    font=\footnotesize,
    rotate=15,
    anchor=north east,
    xshift=10pt,
    align=center,
  }
]

\addplot coordinates {
  (Definite,57.6)
  (Possible,23.3)
  (No/Other,19.1)
};

\end{axis}
\end{tikzpicture}
\caption{Participant interest in GenAI training (N=330).}
\label{fig:genai_training_interest}
\end{figure}
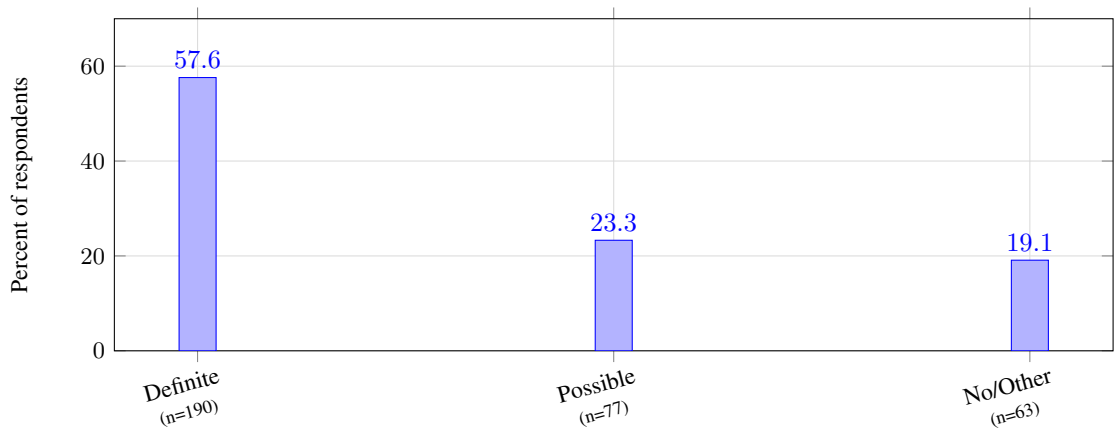

As shown in Figure~\ref{fig:genai_training_interest}, of the 330 participants, 190 (57.6\%) indicated definite interest in GenAI training and 77 (23.3\%) indicated possible interest; topic selections are reported for this combined group (N = 267), with multiple selections allowed.

The most popular choice was \emph{“Using Artificial Intelligence in My Specialized Field”}, selected by nearly half of the participants (49.1\%). This suggests a strong desire for domain-specific training tailored to respondents’ professions or academic backgrounds, rather than general-purpose instruction. The second most selected topic was \emph{“Privacy Protection and Data Preservation”} (47.0\%), reflecting widespread concerns about responsible data use, trust, and security in GenAI systems.

Foundational topics also received substantial interest. Specifically, \emph{“Fundamentals of Generative Artificial Intelligence”} was selected by 38.2\% of participants, while 30.3\% expressed interest in learning how to evaluate model outputs through \emph{“Assessment of Output Quality”}. These results highlight a broad interest in both applied and conceptual training, spanning hands-on usage, ethical considerations, and theoretical foundations.

\vspace{0.5em}
To explore how interests overlap, we conducted a co-selection analysis using \emph{Jaccard similarity}~\cite{travieso2023jaccard}, a measure that captures the proportion of shared selections between two topics. The Jaccard index is calculated as the size of the intersection divided by the size of the union between two sets. This analysis helped reveal clusters of training topics that were frequently selected together.

Table~\ref{tab:training_pairs} shows the most frequently co-selected GenAI training topics and their Jaccard similarities. Several topic pairs exhibit strong conceptual alignment. One of the strongest relationships occurs between \emph{"Using AI in a specialized professional field"} and \emph{"Foundations of generative AI"} (Jaccard = 0.433), suggesting that many participants who seek domain-specific AI applications also want grounding in fundamental GenAI concepts.

\begin{table}[t]
\centering
\caption{Training topics of interest among participants. Ranked by selection frequency. (\(N=267\))}
\label{tab:training_overview}
\begin{tabular}{@{}rlcc@{}}
\toprule
\textbf{Rank} & \textbf{Topic} & \textbf{n Selected} & \textbf{\% of N} \\
\midrule
1 & Using AI in My Specialized Field & 162 & 49.1\% \\
2 & Privacy Protection and Data Preservation & 155 & 47.0\% \\
3 & Fundamentals of GenAI & 126 & 38.2\% \\
4 & Assessment of Output Quality and Credibility & 100 & 30.3\% \\
5 & Writing Prompts Effectively & 100 & 30.3\% \\
6 & Advanced Training on Tools like Copilot and Cursor & 90 & 27.3\% \\
7 & Ethical and Legal Aspects of AI & 82 & 24.9\% \\
\bottomrule
\end{tabular}
\end{table}

\begin{table}[t]
\centering
\caption{Top co-selected training topic pairs. Includes joint count and Jaccard similarity. (\(N=267\))}
\label{tab:training_pairs}
\begin{tabular}{@{}llcc@{}}
\toprule
\textbf{Topic Pair} &  & \textbf{n Both} & \textbf{Jaccard} \\
\midrule
Privacy + Specialized Field & & 95 & 0.428 \\
Specialized Field + Fundamentals & & 87 & 0.433 \\
Privacy + Fundamentals & & 79 & 0.391 \\
Output Quality + Specialized Field & & 78 & 0.424 \\
Prompt Writing + Specialized Field & & 75 & 0.401 \\
Privacy + Ethical/Legal Aspects & & 69 & 0.411 \\
Specialized Field + Copilot / Cursor & & 67 & 0.362 \\
Privacy + Output Quality & & 66 & 0.349 \\
\bottomrule
\end{tabular}
\end{table}

Privacy-related topics also show consistently high co-selection patterns. \emph{“Privacy protection and data preservation”} pairs strongly with both \emph{“Ethical and legal aspects of AI”} (Jaccard = 0.411) and \emph{“Using AI in a specialized field”} (Jaccard = 0.428). This indicates that participants who express interest in AI privacy training also tend to value responsible-use frameworks and applied field competencies.

Skill-building topics cluster heavily with specialized field applications as well. For example, \emph{“Evaluating output quality”} (Jaccard = 0.424) and \emph{“Writing effective prompts”} (Jaccard = 0.401) both frequently co-occur with specialized field training, implying that participants see these technical skills as essential to effective AI use in their professions.

Overall, the co-selection patterns reveal several coherent themes: interest in foundational knowledge combined with applied practice, a strong emphasis on responsible and ethical AI use, and demand for practical skill-building topics. These patterns can guide the design of integrated, bundled training curricula that reflect how learners naturally group their needs.

\begin{tcolorbox}[colback=blue!5!white, colframe=blue!75!black,
title=Summary: Training Interests and Co-Selection Patterns, fonttitle=\bfseries]

A majority of participants expressed interest in GenAI training, with 57.6\% indicating definite interest and an additional 23.3\% reporting possible interest. Nearly half of participants were interested in domain-specific GenAI training, followed closely by privacy and foundational knowledge topics. Co-selection analysis using Jaccard similarity revealed clear thematic clusters. Strong connections emerged between foundational concepts and applied professional use, as well as between privacy and legal-ethical concerns, with several pairs showing moderate to high overlap. Practical skill-building topics such as evaluating output quality and writing effective prompts also clustered closely with specialized field training. These findings suggest that future training programs could be more effective by bundling highly associated topics to match learner interests.

\end{tcolorbox}

\subsection{RQ5 – Barriers, Risks, and Trust in GenAI}

\begin{table}[t]
\centering
\caption{Significant group differences in barriers to GenAI use (one-way ANOVA). (\(N=315\))}
\label{tab:barriers_diffs}
\begin{tabular}{llrrr}
\toprule
\textbf{Measure} & \textbf{Grouping} & \textbf{$F$} & \textbf{$p$} & \textbf{$\eta^2$} \\
\midrule
Lack of trust             & Age group         & 3.652 & 0.0016 & 0.064 \\
Lack of trust             & Sector            & 2.494 & 0.0122 & 0.059 \\
No clear use case         & Age group         & 2.214 & 0.0415 & 0.040 \\
Difficulty learning/using & Age group         & 2.433 & 0.0258 & 0.043 \\
Difficulty learning/using & Employment status & 3.158 & 0.0144 & 0.037 \\
Technical constraints     & Sector            & 1.974 & 0.0492 & 0.047 \\
\bottomrule
\end{tabular}
\end{table}

\begin{table}[t]
\centering
\caption{Significant group differences in concerns about GenAI (one-way ANOVA). (\(N=330\))}
\label{tab:concerns_diffs}
\begin{tabular}{llrrr}
\toprule
\textbf{Measure} & \textbf{Grouping} & \textbf{$F$} & \textbf{$p$} & \textbf{$\eta^2$} \\
\midrule
Job loss               & Sector            & 1.984 & 0.0479 & 0.047 \\
Excessive reliance      & Age group         & 2.185 & 0.0441 & 0.039 \\
Excessive reliance      & Employment status & 3.431 & 0.0091 & 0.041 \\
Content accuracy issues & Age group         & 2.481 & 0.0232 & 0.044 \\
Content accuracy issues & Education level   & 3.600 & 0.0069 & 0.042 \\
Content accuracy issues & Employment status & 4.926 & 0.0007 & 0.057 \\
\bottomrule
\end{tabular}
\end{table}

To move beyond the perceived impact of GenAI, it is crucial to examine the concerns, barriers, and behavioral responses that shape its adoption across different contexts. Although aggregate statistics capture broad patterns of use, they often obscure important variation driven by demographic and institutional factors. In Saudi Arabia, where digital transformation is accelerating and the population is young and highly connected, these dynamics make it essential to examine how barriers and risks vary across demographic groups. Using one-way ANOVA tests, we identify significant variations in barriers to use, concerns about outcomes, and willingness to share personal data across age, education, employment status, and sector. Tables \ref{tab:barriers_diffs} and \ref{tab:concerns_diffs} present these results, highlighting the factors that most strongly differentiate concerns and barrier-related behaviors.

The $F$-value reflects the ratio of variance between groups to variance within groups: higher values indicate stronger group differences relative to within-group variability. The $p$-value represents the probability that the observed differences occurred by chance, with values below $0.05$ conventionally considered statistically significant. The effect size $\eta^2$ quantifies the proportion of variance in the dependent variable explained by group membership.

Table \ref{tab:barriers_diffs} demonstrates that barriers to GenAI adoption are not uniform but vary systematically across demographic and institutional dimensions, including trust, perceived usefulness, usability, and infrastructure. Among these, trust stands out as the most consistent barrier, showing significant differences across both age groups ($F=3.65, p=0.0016, \eta^2=0.064$) and sectors ($F=2.49, p=0.0122, \eta^2=0.059$). Unlike other barriers, it is shaped simultaneously by  contexts, highlighting that confidence in GenAI is unevenly distributed across society.

\begin{figure*}[h!]
	\centering
    
\includegraphics[width=0.6\textwidth, ]{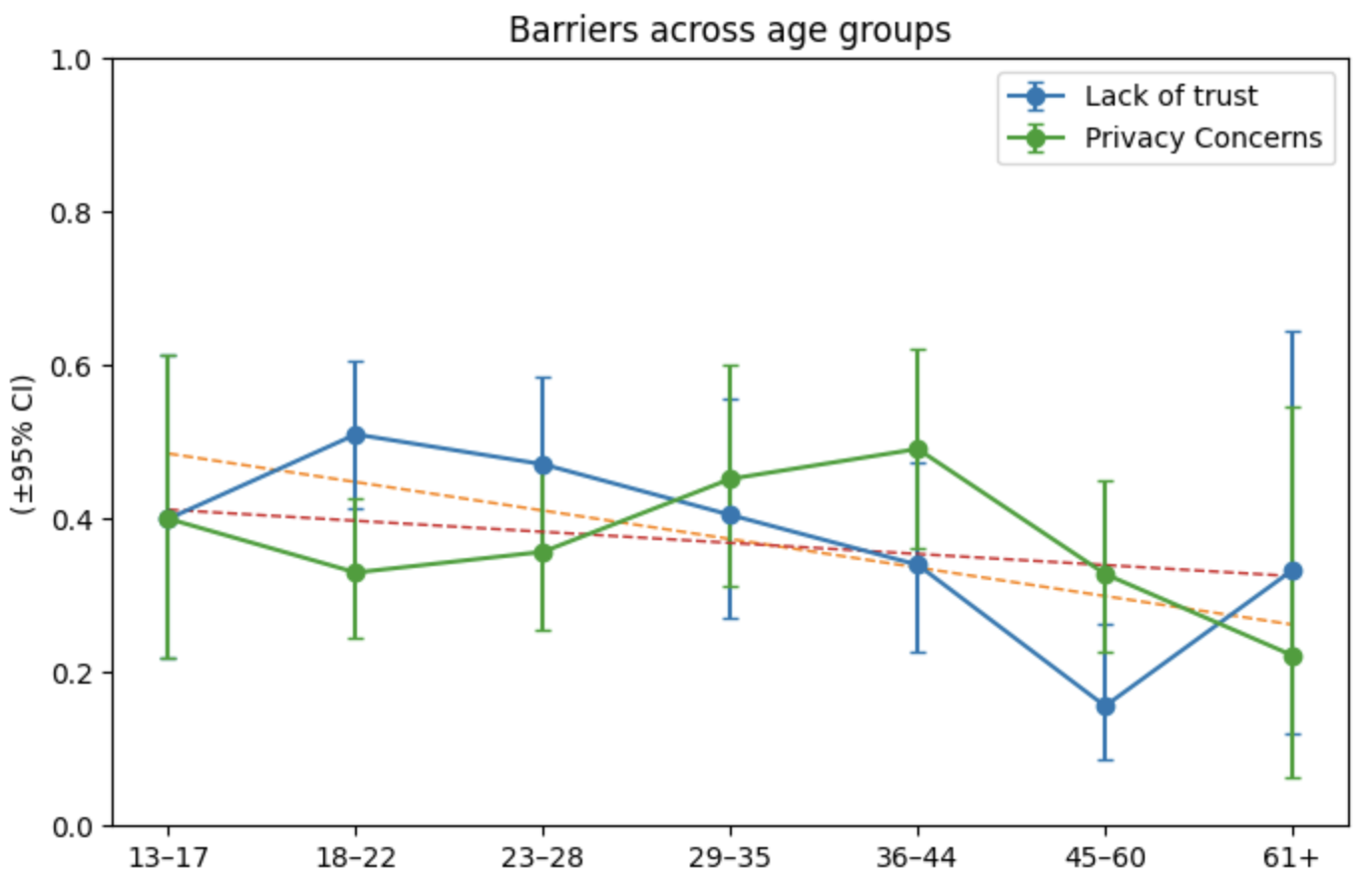}	
	\caption{Barriers to GenAI adoption across age groups. Mean scores (±95\% CI) are shown for lack of trust and privacy concerns.  (\(N=330\))} 
	\label{rii}%
\end{figure*}

Other barriers likewise exhibit meaningful variation across groups. Perceptions of “no clear use case” differ significantly by age ($F=2.21, p=0.0415, \eta^2=0.040$), while difficulties in learning and using GenAI vary across both age categories ($F=2.43, p=0.0258, \eta^2=0.043$) and employment status ($F=3.16, p=0.0144, \eta^2=0.037$). Technical constraints also differ across sectors ($F=1.97, p=0.0492, \eta^2=0.047$), reflecting uneven access to infrastructure and institutional support. Although the effect sizes are generally small to medium, their recurrence across multiple barriers indicating that both demographic and organizational contexts systematically shape how individuals perceive obstacles to adopting GenAI.

To further illustrate these patterns, Figure \ref{rii} compares trust and privacy concerns across age groups. Privacy concerns peak among participants aged 36–44 (mean = 0.49), suggesting heightened sensitivity to data risks during mid-career stages, whereas both younger cohorts (13–17 and 18–22) and older adults (61+) report substantially lower levels of concern. Trust shows an inverse pattern: skepticism is highest among young adults aged 18–22 (mean = 0.51) but decreases steadily with age, reaching its lowest level among those aged 45–60 (mean = 0.16). These findings reveal a generational divide in how risks are perceived, mid-career professionals emphasize privacy threats, while younger adults express stronger doubts about the reliability of GenAI.


Table \ref{tab:concerns_diffs} shows that concerns about GenAI are significantly differentiated across demographic contexts. Fears of job loss vary significantly by sector ($F=1.98, p=0.0479, \eta^2=0.047$), pointing to uneven perceptions of labor market risks across industries. Doubts about excessive reliance on AI differ both by age ($F=2.19, p=0.0441, \eta^2=0.039$) and employment status ($F=3.43, p=0.0091, \eta^2=0.041$), emphasizing how life stage and professional role shape sensitivity to technological dependence. Most notably, concerns about content accuracy, such as false or misleading information, emerge as the most broadly distributed, showing significant variation across age ($F=2.48, p=0.0232, \eta^2=0.044$), education ($F=3.60, p=0.0069, \eta^2=0.042$), and employment status ($F=4.93, p=0.0007, \eta^2=0.057$). This breadth of differentiation indicates that anxieties around misinformation and reliability reaches across generational, educational, and occupational boundaries, constituting their most widespread concern. Although effect sizes remain in the small-to-medium range, their consistency highlights that concerns about GenAI are not uniform but systematically shaped by different context.

\begin{figure*}[h!]
	\centering
\includegraphics[width=0.8\textwidth, ]{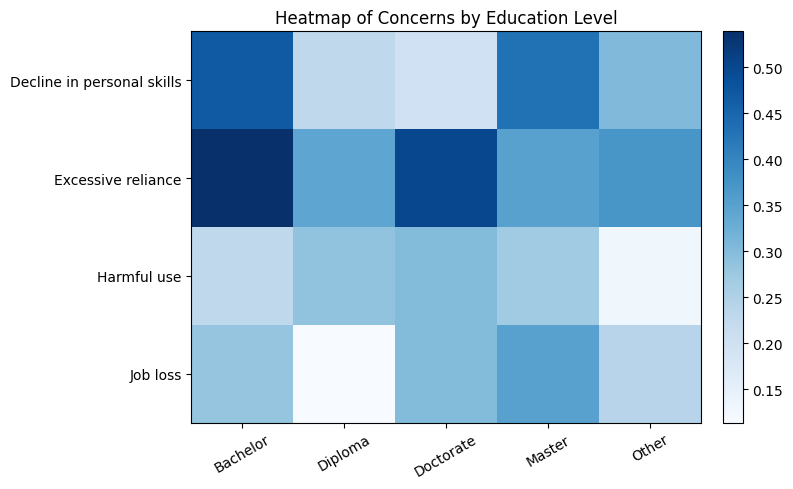}	
	\caption{Heatmap of concern levels across education levels.  (\(N=330\))} 
	\label{heat_m}%
\end{figure*}

The heatmap in Figure \ref{heat_m} demonstrates that concerns about GenAI vary systematically by education level, revealing distinct and consequential patterns. Concerns about excessive reliance are most prominent among Bachelor’s and Doctorate holders (means $\approx$ 0.54 and 0.50), indicating that both broadly trained and research-intensive groups are particularly sensitive to the risks of overdependence on AI tools. Concerns about a decline in personal skills are likewise elevated among Bachelor’s (0.47) and Master’s (0.43) respondents, indicating that academically engaged populations perceive AI as a potential threat to core competencies such as writing and analytical reasoning. By contrast, fears of job loss remain relatively modest overall but peak among Master’s graduates (0.35), reflecting anxieties about career vulnerability in mid- to advanced professional stages. Finally, perceived risks of harmful use are consistently low across all educational groups, with respondents in the ``Other'' category reporting the lowest concern (0.13). Collectively, these results show that educational background not only influences the intensity but also the nature of AI-related concerns, whether employment security or technological dependence, emphasizing the need for governance and awareness strategies tailored to specific educational levels.


\begin{tcolorbox}[colback=blue!5!white, colframe=blue!75!black,
title={Summary: Barriers, Risks, and Trust in GenAI}, fonttitle=\bfseries]

Barriers and concerns about GenAI vary systematically across age, education, employment status, and sector rather than being uniform. Trust is the most consistent barrier, with additional obstacles such as unclear use cases, usability challenges, and technical constraints reflecting unequal readiness and support. Privacy and misinformation worries are especially salient, mid-career adults emphasize data risks, younger users show stronger skepticism toward AI outputs, and university-educated groups are more concerned about overreliance, skill decline, and job loss, highlighting the need for targeted rather than one-size-fits-all interventions.

\end{tcolorbox}

\subsection{RQ6 – Willingness to Share Personal Data by User Status and Data Type}

\begin{figure*}[h!]
	\centering
\includegraphics[width=0.7\textwidth, ]{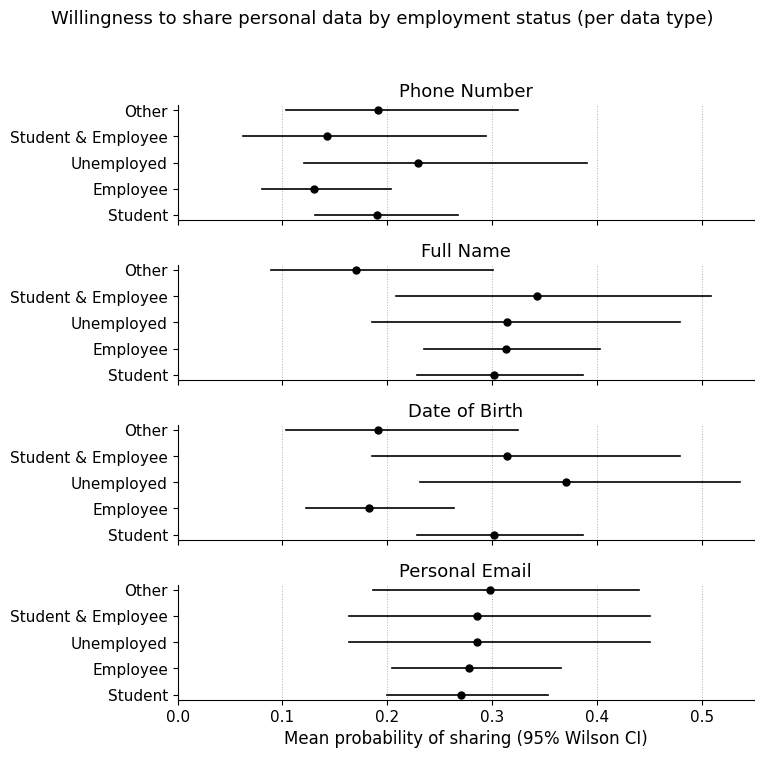}	
	\caption{Mean probability of sharing personal data with GenAI tools by employment status and data type.  (\(N=330\))} 
	\label{share}%
\end{figure*}


Figure \ref{share} shows how willingness to share personal data with GenAI tools varies by employment status and data type. Students and unemployed respondents consistently report the highest probabilities of disclosure, including for sensitive attributes such as date of birth. For example, unemployed respondents have a mean probability of sharing their date of birth of about 0.37, compared to roughly 0.18 among employees. This gap suggests that employment context shapes both perceived necessity and perceived risk when interacting with GenAI tools. In contrast, employees report the lowest sharing rates overall, reflecting stronger risk awareness and stricter norms in formal work settings. Respondents who combine study and employment fall in between, balancing exposure to GenAI with professional caution. Across all groups, personal email is the most frequently shared data type, while phone numbers are shared the least, indicating a clear hierarchy of perceived sensitivity. Together, these patterns show that data-sharing behaviors are structured by employment status, with implications for targeted awareness efforts and institutional safeguards.

\begin{tcolorbox}[colback=blue!5!white, colframe=blue!75!black,
title=Summary: Data Sharing and User Status, fonttitle=\bfseries]

Willingness to share personal data with GenAI tools varies by employment status. Students and unemployed respondents are most likely to disclose sensitive details, while employees share the least, indicating stronger risk awareness in formal work settings. Across all groups, personal email is shared most and phone numbers least, revealing a clear hierarchy of perceived sensitivity.

\end{tcolorbox}

\subsection{RQ7 – Future Expectations, Policy Preferences, and Personal Reflections}

In this section, we analyze participants’ responses to gain deeper insight into their personal views, concerns, and aspirations regarding GenAI. Alongside structured survey items, participants also provided general sentiment ratings and open-ended reflections. These combined sources offer a rich perspective on how people in Saudi Arabia experience and interpret GenAI in everyday life. Through thematic coding~\cite{braun2006using} and qualitative analysis, we identified recurring patterns in sentiment, proposed solutions, and culturally grounded values. These reflections complement the quantitative results and highlight meaningful areas for future development, policy design, and public engagement.

\subsubsection{User Impressions of GenAI}

Before analyzing open-ended comments, participants were asked to select the statement that best described their overall impression of GenAI tools. As shown in Table \ref{tab:overall_attitudes}, nearly two-thirds of respondents (64.8\%) expressed generally positive attitudes, including 40.6\% who viewed GenAI as a useful tool offering tangible benefits and 24.2\% who described it as highly transformative. Approximately one-third (30.9\%) reported a neutral stance, indicating a balanced perception of benefits and risks. Only a small minority (4.2\%) expressed negative views.

These sentiment ratings provided a structured baseline for understanding participants’ subsequent free-text reflections.

\begin{table}[htbp]
\centering
\caption{Overall attitudes toward GenAI tools. (\(N = 330\))}
\label{tab:overall_attitudes}
\begin{tabular}{lrr}
\toprule
Response option & n & \% of \(N\) \\
\midrule
Positive -- a useful tool with tangible benefits          & 134 & 40.6\% \\
Very positive -- a transformative, game-changing technology & 80  & 24.2\% \\
Neutral -- balanced advantages and disadvantages           & 102 & 30.9\% \\
Negative -- risks may outweigh benefits                    & 6   & 1.8\% \\
Very negative -- harms greatly outweigh benefits           & 8   & 2.4\% \\
\midrule
Total                                                      & 330 & 100\% \\
\bottomrule
\end{tabular}
\end{table}

Following this quantitative snapshot, we analyzed 330 responses in which participants described their general impressions of GenAI tools. Thematic coding revealed that the majority expressed a positive tone, frequently highlighting benefits such as speed, ease of use, and productivity. About one in six participants shared explicitly favorable opinions, as reflected in comments like: \textit{``Excellent when used in moderation and balance.''}

The most common theme was \textit{time-saving}, with many users noting how GenAI helps complete tasks more quickly. One participant stated: \textit{``Saves a lot of time, especially at work or when we need ideas quickly.''} Other commonly cited benefits included \textit{better comprehension}, \textit{summarization}, and \textit{creative idea support}.

At the same time, a sizable portion of users expressed caution. Some worried about \textit{overreliance}, fearing that excessive use might weaken critical thinking or creativity. Others mentioned \textit{inaccuracy} and the risk of system failure or discontinuation. Concerns about \textit{privacy}, \textit{job loss}, and \textit{misinformation} were less frequent but present.

A number of participants also offered reflective views on appropriate usage. These comments emphasized the importance of using GenAI in moderation, verifying outputs, and treating it as a tool to support, not replace human effort. Overall, user sentiment was cautiously optimistic: people are excited about GenAI’s potential but want to see responsible, thoughtful use.

\begin{tcolorbox}[colback=blue!5!white, colframe=blue!75!black,
title=Summary: User Impressions, fonttitle=\bfseries]
Most participants expressed positive views of GenAI, noting its speed, usefulness, and productivity benefits. Time-saving and improved comprehension were common themes. However, some participants voiced concerns about overreliance, accuracy, and ethical risks. Overall, sentiment was cautiously optimistic, favoring responsible, balanced, and well-informed use.
\end{tcolorbox}

\subsubsection{Desired Initiatives and Solutions}

Participants also proposed a variety of ideas for improving GenAI adoption and integration in Saudi Arabia. The most common request was for \textit{education and training}, spanning school and university use. Respondents envisioned GenAI tools that help students understand material more easily or support educators. One user wrote: \textit{``I hope to see initiatives that focus on using artificial intelligence in education and healthcare, like tools that help students understand more easily.''}

Another recurring theme was the need for \textit{reliable and trustworthy systems}. Many participants called for GenAI tools that provide verifiable outputs, reduce hallucinations, and protect user privacy. Concerns about \textit{data security} were frequently tied to calls for safe, transparent systems.

Specific domains of interest included \textit{healthcare}, especially tools for elderly care and chronic disease management. Other participants mentioned the potential for GenAI in \textit{government services}, \textit{urban planning}, and \textit{social programs}.

Participants also called for more \textit{inclusive and accessible} tools, including \textit{free or low-cost services}, support for \textit{people with disabilities or limited resources}, and tools designed for the \textit{elderly}. One person noted: \textit{``These innovations should make tasks easier for the elderly and individuals with disabilities.''}

A smaller group focused on technical and policy-related solutions. They called for \textit{all-in-one assistants}, tools that work offline or locally (in line with Saudi data policies), and stronger support for \textit{Arabic language and cultural context}. Several tied their proposals to \textit{Vision 2030}, suggesting that GenAI tools should align with national goals.

\begin{tcolorbox}[colback=blue!5!white, colframe=blue!75!black,
title=Summary: Desired Solutions and Priorities, fonttitle=\bfseries]
Participants called for GenAI initiatives that emphasize education, reliability, and privacy. Popular suggestions included tools for schools, healthcare, and public services. Accessibility, Arabic language support, and cultural alignment were key priorities. Many viewed GenAI as a tool for national development, not just innovation.
\end{tcolorbox}

\subsubsection{Overall Participant Reflections}

In their final comments, participants were invited to share closing thoughts on GenAI or the survey. A total of 34 valid responses offered a mix of excitement, caution, and hope for the future.

The most frequently mentioned concern was \textit{privacy and accuracy}. Several participants emphasized the need for \textit{fact-checked, secure, and reliable systems}. One user wrote: \textit{``AI is evolving rapidly… but it can be a bit scary. I hope people become more aware of how to use it in a way that benefits us without compromising our privacy.''}

Cultural and linguistic identity also featured prominently. Respondents called for tools that reflect \textit{local values}, support \textit{Arabic content}, and respect \textit{Islamic principles}. One participant said: \textit{``There is a lack of quality Arabic content and local culture in these tools. I hope to see strong tools developed in Saudi Arabia.''}

Several participants connected GenAI development to \textit{national goals} like Vision 2030. These reflections showed a sense of civic duty and pride, urging developers to build technologies that advance Saudi Arabia’s future. As one user noted: \textit{``I urge you to take on the responsibility and trust to serve your great country, in line with the vision of SA.''}

A few participants expressed curiosity about more advanced topics like \textit{open-source models} and \textit{AI agents}, suggesting an emerging public interest in deeper technical concepts.

\begin{tcolorbox}[colback=blue!5!white, colframe=blue!75!black,
title=Summary: Final Reflections, fonttitle=\bfseries]
Final reflections emphasized privacy, Arabic support, and national alignment. Many participants want GenAI to reflect Saudi values and contribute to Vision 2030. Others showed growing curiosity about technical topics. Overall, these reflections highlight public interest in ethical, local, and impactful GenAI.
\end{tcolorbox}

\section{Limitations}
\label{sec:Limitations}


While this study provides one of the first empirical investigations into the awareness, adoption, and trust of Generative Artificial Intelligence (GenAI) within the Saudi Arabian context, several limitations must be acknowledged. First, although the survey included 330 participants representing diverse regions, age groups, and professional sectors, the sample may not fully reflect the demographic diversity of the Kingdom. Certain groups, such as older adults, individuals from rural areas, or those with limited digital access may have been underrepresented.

Additionally, the study relies primarily on self-reported data, which may be subject to recall bias or social desirability effects. Respondents might have overestimated their familiarity or engagement with GenAI tools, particularly given the growing public enthusiasm surrounding emerging technologies. While self-report surveys are useful for gauging perceptions and attitudes, future research could incorporate behavioral or usage-based data to obtain a more objective assessment of GenAI adoption patterns. 

Moreover, the temporal context of the study represents another limitation. Data collection occurred during a period of rapid technological and policy development in Saudi Arabia, coinciding with major national initiatives. As these initiatives mature and new regulations or models emerge, public perceptions of GenAI may evolve substantially. Follow-up studies would therefore be valuable in tracking these dynamic changes over time. Finally, the findings are context-specific to Saudi Arabia and may not be directly transferable to other Gulf or Arabic-speaking countries, where variations in regulatory frameworks, cultural values, and levels of digital infrastructure could yield different outcomes.
\section{Conclusion}
\label{sec:Conclusion}

This study provides an early survey-based snapshot of GenAI engagement in Saudi Arabia within the context of the Kingdom’s Vision~2030 transformation. Drawing on responses from 330 participants across regions, age groups, and sectors, we examined awareness, adoption patterns, perceived impacts, training needs, risk perceptions, and future expectations. Results indicate that GenAI use is  widespread, particularly among younger adults, with most respondents relying on it for text-oriented and informational tasks such as research assistance, writing, presentations, email drafting, brainstorming, and summarization, while more advanced uses (e.g., coding or multimodal creation) remain limited. Trust in AI outputs is cautious: respondents report frequently reviewing generated content and express notable concerns around data misuse and potential job displacement. While most avoid sharing highly sensitive data, some disclose personal information, reflecting enthusiasm tempered by risk awareness.

\section*{Acknowledgements}
\label{sec:Acknowledgements}

The authors would like to acknowledge Abdulwahed AlAbdulwahed and Kadhem Almarri for their assistance and participation in this research.


\bibliographystyle{IEEEtran}
\bibliography{references}

\begin{thebibliography}{10}
\providecommand{\url}[1]{#1}
\csname url@samestyle\endcsname
\providecommand{\newblock}{\relax}
\providecommand{\bibinfo}[2]{#2}
\providecommand{\BIBentrySTDinterwordspacing}{\spaceskip=0pt\relax}
\providecommand{\BIBentryALTinterwordstretchfactor}{4}
\providecommand{\BIBentryALTinterwordspacing}{\spaceskip=\fontdimen2\font plus
\BIBentryALTinterwordstretchfactor\fontdimen3\font minus \fontdimen4\font\relax}
\providecommand{\BIBforeignlanguage}[2]{{%
\expandafter\ifx\csname l@#1\endcsname\relax
\typeout{** WARNING: IEEEtran.bst: No hyphenation pattern has been}%
\typeout{** loaded for the language `#1'. Using the pattern for}%
\typeout{** the default language instead.}%
\else
\language=\csname l@#1\endcsname
\fi
#2}}
\providecommand{\BIBdecl}{\relax}
\BIBdecl

\bibitem{sdaia_report}
\BIBentryALTinterwordspacing
{Saudi Data and Artificial Intelligence Authority (SDAIA)}, ``Generative artificial intelligence: Promising prospects for a better future,'' 2024, accessed: 2025-09-14. [Online]. Available: \url{https://www.saudigazette.com.sa/article/652863}
\BIBentrySTDinterwordspacing

\bibitem{sdaia_guidelines}
\BIBentryALTinterwordspacing
------, ``Sdaia publishes generative ai guidelines,'' 2024, accessed: 2025-09-14. [Online]. Available: \url{https://digitalpolicyalert.org/event/17196-published-sdaia-generative-ai-guidelines}
\BIBentrySTDinterwordspacing

\bibitem{suguna2021artificial}
S.~K. Suguna, M.~Dhivya, and S.~Paiva, ``Artificial intelligence (ai): Recent trends and applications,'' 2021.

\bibitem{bick-2024}
\BIBentryALTinterwordspacing
A.~Bick, A.~Blandin, and D.~Deming, ``{The rapid adoption of generative AI},'' \emph{SSRN Electronic Journal}, 1 2024. [Online]. Available: \url{https://doi.org/10.2139/ssrn.4964384}
\BIBentrySTDinterwordspacing

\bibitem{freeman-2025}
\BIBentryALTinterwordspacing
J.~Freeman, ``{Student Generative AI Survey 2025},'' \emph{Higher Education Policy Institute}, 2 2025. [Online]. Available: \url{https://www.hepi.ac.uk/reports/student-generative-ai-survey-2025/}
\BIBentrySTDinterwordspacing

\bibitem{ravselj-2025}
D.~Rav{\v{s}}elj, D.~Ker{\v{z}}i{\v{c}}, N.~Toma{\v{z}}evi{\v{c}}, L.~Umek, N.~Brezovar, N.~A. Iahad, A.~A. Abdulla, A.~Akopyan, M.~W.~A. Segura, J.~AlHumaid \emph{et~al.}, ``Higher education students’ perceptions of chatgpt: A global study of early reactions,'' \emph{PLoS One}, vol.~20, no.~2, p. e0315011, 2025.

\bibitem{davis2024early}
J.~P. Davis and J.~B. Li, ``Early adoption of generative ai by global business leaders: Insights from an insead alumni survey,'' \emph{arXiv preprint arXiv:2404.04543}, 2024.

\bibitem{syed2024awareness}
W.~Syed, A.~Bashatah, K.~Alharbi, S.~S. Bakarman, S.~Asiri, and N.~Alqahtani, ``Awareness and perceptions of chatgpt among academics and research professionals in riyadh, saudi arabia: Implications for responsible ai use,'' \emph{Medical Science Monitor: International Medical Journal of Experimental and Clinical Research}, vol.~30, pp. e944\,993--1, 2024.

\bibitem{baek-2024}
\BIBentryALTinterwordspacing
C.~Baek, T.~Tate, and M.~Warschauer, ``{“ChatGPT Seems Too Good to be True”: College Students’ Use and Perceptions of Generative AI},'' \emph{Computers and Education Artificial Intelligence}, p. 100294, 9 2024. [Online]. Available: \url{https://doi.org/10.1016/j.caeai.2024.100294}
\BIBentrySTDinterwordspacing

\bibitem{pew-2025}
\BIBentryALTinterwordspacing
C.~McClain, B.~Kennedy, J.~Gottfried, M.~Anderson, and G.~Pasquini, ``{How the U.S. Public and AI Experts View Artificial Intelligence},'' \emph{Pew Research Center}, Apr. 2025. [Online]. Available: \url{https://www.pewresearch.org/internet/2025/04/03/how-the-us-public-and-ai-experts-view-artificial-intelligence/}
\BIBentrySTDinterwordspacing

\bibitem{d2025ai}
S.~D’haeseleer, K.~Van~Damme, H.~Cools, S.~Van~Leuven, and T.~Evens, ``Ai divides in newsrooms? how journalists in the low countries use and perceive generative ai,'' \emph{Journalism Practice}, pp. 1--28, 2025.

\bibitem{wolf-2024}
\BIBentryALTinterwordspacing
V.~Wolf and C.~Maier, ``{ChatGPT usage in everyday life: A motivation-theoretic mixed-methods study},'' \emph{International Journal of Information Management}, vol.~79, p. 102821, 6 2024. [Online]. Available: \url{https://doi.org/10.1016/j.ijinfomgt.2024.102821}
\BIBentrySTDinterwordspacing

\bibitem{dillon-2025}
\BIBentryALTinterwordspacing
E.~W. Dillon, S.~Jaffe, N.~Immorlica, and C.~Stanton, ``{Shifting Work Patterns with Generative AI},'' \emph{SSRN Electronic Journal}, 1 2025. [Online]. Available: \url{https://doi.org/10.2139/ssrn.5250761}
\BIBentrySTDinterwordspacing

\bibitem{lee-2025}
\BIBentryALTinterwordspacing
H.-P.~H. Lee, A.~Sarkar, L.~Tankelevitch, I.~Drosos, S.~Rintel, R.~Banks, and N.~Wilson, ``The impact of generative ai on critical thinking: Self-reported reductions in cognitive effort and confidence effects from a survey of knowledge workers,'' in \emph{Proceedings of the 2025 CHI Conference on Human Factors in Computing Systems}, ser. CHI '25.\hskip 1em plus 0.5em minus 0.4em\relax New York, NY, USA: Association for Computing Machinery, 2025. [Online]. Available: \url{https://doi.org/10.1145/3706598.3713778}
\BIBentrySTDinterwordspacing

\bibitem{alammari-2024}
\BIBentryALTinterwordspacing
A.~Alammari, ``{Evaluating generative AI integration in Saudi Arabian education: a mixed-methods study},'' \emph{PeerJ Computer Science}, vol.~10, p. e1879, 2 2024. [Online]. Available: \url{https://doi.org/10.7717/peerj-cs.1879}
\BIBentrySTDinterwordspacing

\bibitem{ma2025dancingbearcolleaguesharpened}
\BIBentryALTinterwordspacing
R.~Ma, M.~Dedema, and A.~Cox, ``A dancing bear, a colleague, or a sharpened toolbox? the cautious adoption of generative ai technologies in digital humanities research,'' 2025. [Online]. Available: \url{https://arxiv.org/abs/2404.12458}
\BIBentrySTDinterwordspacing

\bibitem{cronbach1951coefficient}
L.~J. Cronbach, ``Coefficient alpha and the internal structure of tests,'' \emph{psychometrika}, vol.~16, no.~3, pp. 297--334, 1951.

\bibitem{spearman1961proof}
C.~Spearman, ``The proof and measurement of association between two things.'' 1961.

\bibitem{kruskal1952use}
W.~H. Kruskal and W.~A. Wallis, ``Use of ranks in one-criterion variance analysis,'' \emph{Journal of the American statistical Association}, vol.~47, no. 260, pp. 583--621, 1952.

\bibitem{tomczak2014need}
M.~Tomczak and E.~Tomczak, ``The need to report effect size estimates revisited. an overview of some recommended measures of effect size,'' 2014.

\bibitem{montgomery2021introduction}
D.~C. Montgomery, E.~A. Peck, and G.~G. Vining, \emph{Introduction to linear regression analysis}.\hskip 1em plus 0.5em minus 0.4em\relax John Wiley \& Sons, 2021.

\bibitem{mackinnon1985some}
J.~G. MacKinnon and H.~White, ``Some heteroskedasticity-consistent covariance matrix estimators with improved finite sample properties,'' \emph{Journal of econometrics}, vol.~29, no.~3, pp. 305--325, 1985.

\bibitem{fisher1970statistical}
R.~A. Fisher, ``Statistical methods for research workers,'' in \emph{Breakthroughs in statistics: Methodology and distribution}.\hskip 1em plus 0.5em minus 0.4em\relax Springer, 1970, pp. 66--70.

\bibitem{travieso2023jaccard}
G.~Travieso and L.~d.~F. Costa, ``The jaccard similarity mean,'' \emph{arXiv preprint arXiv:2311.12959}, 2023.

\bibitem{braun2006using}
V.~Braun and V.~Clarke, ``Using thematic analysis in psychology,'' \emph{Qualitative research in psychology}, vol.~3, no.~2, pp. 77--101, 2006.

\end{thebibliography}
\end{document}